\documentclass[aps,prd,amsmath,floats,floatfix,
  superscriptaddress,
  nofootinbib,
  notitlepage,
  showpacs]{revtex4-1}
  
\usepackage{ulem}
\usepackage{sidecap}
\usepackage{amssymb}
\usepackage{amsmath}
\usepackage{verbatim}
\usepackage{mathrsfs}
\usepackage{amsfonts}
\usepackage{latexsym}
\usepackage{epsfig}
\usepackage{color}
\usepackage{graphicx}
\usepackage{units}
\usepackage{overpic}
\usepackage[hidelinks]{hyperref}
\usepackage{afterpage}
\usepackage{float}
\usepackage{mathtools}
\usepackage{xcolor}
\hypersetup{
    colorlinks,
    linkcolor={red!50!black},
    citecolor={blue!50!black},
    urlcolor={blue!80!black}
}

\usepackage{booktabs, multirow} 
\usepackage{soul}
\usepackage{changepage,threeparttable} 
\usepackage{CJKutf8}

\begin{document}


\definecolor{orange}{rgb}{0.9,0.45,0} 
\definecolor{applegreen}{rgb}{0.055, 0.591, 0.0530}
\newcommand{\vjp}[1]{{\textcolor{blue}{[VJ: #1]}}}
\newcommand{\mariana}[1]{{\textcolor{brown}{[mariana: #1]}}}
\newcommand{\dario}[1]{{\textcolor{red}{[Dario: #1]}}}
\newcommand{\dmc}[1]{{\textcolor{applegreen}{[DM: #1]}}}


\title{
Electric traversable wormhole supported by a charged scalar field
}

\author{V\'ictor Jaramillo }
\affiliation{Instituto de Ciencias Nucleares, Universidad Nacional
  Aut\'onoma de M\'exico, Circuito Exterior C.U., A.P. 70-543,
  M\'exico D.F. 04510, M\'exico}
\affiliation{Department of Astronomy, University of Science and Technology of China, Hefei, 230026, China}
  
\author{Mariana Lira}
\affiliation{Instituto de Ciencias Nucleares, Universidad Nacional
  Aut\'onoma de M\'exico, Circuito Exterior C.U., A.P. 70-543,
  M\'exico D.F. 04510, M\'exico}

\author{Daniel Mart\'inez-Carbajal}
\affiliation{Instituto de Ciencias Nucleares, Universidad Nacional
  Aut\'onoma de M\'exico, Circuito Exterior C.U., A.P. 70-543,
  M\'exico D.F. 04510, M\'exico}

\author{Dar\'io N\'u\~nez }
\affiliation{Instituto de Ciencias Nucleares, Universidad Nacional
  Aut\'onoma de M\'exico, Circuito Exterior C.U., A.P. 70-543,
  M\'exico D.F. 04510, M\'exico}
\affiliation{Departamento de Matemática da Universidade de Aveiro and Centre for Research and Development in Mathematics and Applications (CIDMA), Campus de Santiago, 3810-183 Aveiro, Portugal.}


\date{\today}

\begin{abstract}
Solving the Einstein-Klein-Gordon-Maxwell system, we construct and analyze the properties of an electrically charged wormhole, formed from a complex, massive scalar field, with self-interaction, and endowed with an electric charge. The scalar field is minimally coupled to the gravitational and the Maxwell field. Covering regions of the value of the different parameters of such wormhole, we present the dependence of the form of the solution with respect to the value of the different parameters, emphasising the role played by the charge in the configurations; we focus on the region for large values of the self-interaction parameter and found a generic behavior of the scalar field, which in turn allows us to determine explicit analytic expressions for the fields, the metric function and the global quantities such as the Komar mass and the particle number.  The motion of charges in these spacetimes is also reported.
\end{abstract}

\maketitle

\section{Introduction}

To see Einstein's paradigm, the geometry and the matter are interrelated, in the case when the matter violates the energy conditions, not only offers a better understanding of such paradigm but also could describe some physical scenario.

Indeed, the cosmological fact that the expansion of the Universe is accelerating \cite{Perlmutter:1999jt,Abdalla:2022yfr} actually demands the existence of matter that violates at least the strong energy condition. The models that support the inflationary period of the Universe, also demand the presence of such type of matter. Moreover, in a recent work \cite{Jaramillo:2023twi}, some of the authors have shown that it is possible to find situations where that kind of matter, actually one of the most extreme examples is phantom matter with negative kinetic terms, can be confined, which might prove to be a step closer to describing a possible way of formation of one of the most intriguing examples of spacetimes with exotic configuration and topology: the wormholes.

The wormhole solutions are {\it bona fide} as they are actually solving the Einstein equations together with a stress energy tensor describing the matter which violates the null energy condition \cite{Morris:1988cz,Morris:1988tu,Bolokhov:2021fil,Lobo:2005us}. It seems likely that the complex scalar field is a type of matter within hard core General Relativity, which might better describe the exotic matter and has been analyzed in several works, as in Refs. \cite{Liebling:2012fv,Matos:1998vk,Urena-Lopez:2019kud,Gao:2023gof, Li:2013nal}, for instance.

It is precisely in the context of the scalar field, where 50 years ago one of the first exact asymptotically flat and everywhere regular wormhole solutions was found \cite{Bronnikov:1973fh, Ellis:1973yv}. Subsequently, some other solutions have been found, such as configurations with self-interacting scalar fields \cite{Dzhunushaliev:2008bq,Chew:2019lsa}, a self-interacting triplet \cite{Jusufi:2018waj} and solutions with multiple scalar fields \cite{Nojiri:2023dvf, Carvente:2019gkd, Hoffmann:2017vkf}.

Several properties of the wormhole have been established, even since Morris and Thorne's pioneer work \cite{Morris:1988cz}, in the usual pseudo-Riemannian manifold and within the General Theory of Relativity. In Ref. \cite{Carvente:2019gkd} was shown that when the matter building up the wormhole is a massive (complex in general) scalar field, it has to have a nonzero self-coupling constant $\lambda$, to obtain mirror symmetry relative to the throat solutions. Also, as is the case for the boson stars \cite{Kaup:1968zz}, the asymptotic flatness requirement demands that the pulsation frequency $\omega$ is bounded by the value of the mass parameter $\mu$ of the scalar field in consideration: $|\omega| \leq \mu$. On the other hand, it has been shown that both, the simplest realizations of wormholes formed by a massless scalar field (by means of linear perturbations and full non-linear numerical evolution analysis \cite{Gonzalez:2008wd,Shinkai:2002gv, Gonzalez:2008xk}) and the complex massive scalar field wormhole with self-interaction (using perturbation theory \cite{Dzhunushaliev:2008bq}) lead to the same conclusion regarding the instability of the configuration. No stable wormhole solution has been presented; there is also not a clear idea of the formation process of a wormhole. The authors in \cite{Chew:2019lsa} demonstrate several facts: that regular wormhole solutions supported by a complex ghost scalar field with a quartic potential, the coefficient being the self-interaction term, $\lambda$, exist for all values $0 \leq \omega \leq \mu$; that the wormhole masses lie within a region enclosed by the curves  $\omega/\mu=0$ and $\omega/\mu=1$ being positive for $\omega/\mu\sim0$ and negative for $\omega/\mu \rightarrow 1$, and that when $\lambda \rightarrow \infty$ the mass increases without limit and the value of the scalar field at the throat, $\phi_c$ tends to zero.

In order to continue with the understanding of such spacetimes, a step forward is to continue exploring the features of wormholes when other fields are present, recently for example, particular wormholes that do not require matter out of the ordinary have been presented and discussed in the framework of the Einstein-Dirac-Maxwell theory \cite{Blazquez-Salcedo:2020czn,Bolokhov:2021fil,Konoplya:2021hsm,Blazquez-Salcedo:2021udn}, and possible ideas on the direction of formation are presented in\cite{Maso-Ferrando:2023wtz}, within $f(R)$ theories. In this manuscript, a regular, spherically symmetric electric wormhole solution is constructed following the procedure to endow with an electric field a boson star \cite{Jetzer:1989av} by coupling the (complex) scalar field to the electromagnetic one by means of a gauge covariant derivative. Some of the authors of the present work used such formulation to construct a magnetized boson star \cite{Jaramillo:2022gcq}. This procedure preserves gauge invariance contrary to the models where the scalar field interacts with the electromagnetic one through a product $e^{\alpha\phi}\,F^2$ directly at the Lagrangian (see e.g., \cite{DelAguila:2015isj,Huang:2019arj}). Analogous wormhole solutions in the Einstein-Klein-Gordon-Maxwell system have been obtained, to the best of our knowledge only in Ref. \cite{Gonzalez:2009hn} for the case of a massless, real scalar field without coupling (other than that of the gravitational interaction) with the electromagnetic field.  

In the present manuscript, we explore the electric wormhole properties as a function of the value of the parameters, with emphasis on the role of the electric charge; we show that the charge does affect the properties of the wormhole, being the most notable that on the asymptotic mass of the spacetime. Our analysis suggested a peculiar behavior of the functions when the self-interaction parameter $\lambda$ is large; following the seminal work of Colpi and collaborators \cite{Colpi:1986ye} regarding boson stars, we were able to determine analytic expressions for the scalar field and for the total mass for large values of $\lambda$, a fact that we corroborate with the actual numerical solutions of the system for such cases. Such expressions allow for a better understanding of the parameters of the system, namely the scalar field mass and frequency, $\mu$, $\omega$ and the electric charge $q$, in determining the total mass and the particle number of the solutions. 

The work is organized as follows: In Section \ref{Sec:Theoretical}, we introduce the model, fix our spacetime time to be static with spherical symmetry, introduce the ans\"atze for the charged complex scalar field, and write down the field equations for the metric coefficients, the scalar and the Maxwell field with the scalar current being the source of the Maxwell field, finally, we present expressions for some global quantities, namely the Komar mass and the total number of particles.

In Section \ref{Sec:Solutions}, we give boundary conditions to obtain  regular,  asymptotically flat spacetimes together with an important constraint among the values of the fields at the wormhole's throat, and describe our code for solving the field equations with the needed conditions. Next, we sweep several intervals of value for the parameters, stressing the role of the electric charge in the configurations obtained. We present the profile of the total mass of the spacetime and the value of the throat radius, ${\cal G}$, as a function of the value of the scalar field at the throat. We also present the total mass as a function  of $\lambda$ for several values of the field frequency $\omega$ and the charge $q$, also present the profile of the electric field and that of the energy density, $\tau$, as a function of the radius.  We present also the plots of the total mass as a function of the frequency. These results suggest a behavior of the scalar field for large values of the interaction parameter, $\lambda$. In section \ref{large_lambda}, we develop such analysis and obtain analytical expressions for the fields and  the global quantities, proving its validity with the actual solutions. In 
section \ref{motion} we analyze the particle motion for spacetimes with positive, negative or zero total mass, for neutral and charged masses. Finally, in Section \ref{Sec:Conclusions} we give our conclusions. Throughout this work, we use units with $c=G=1$ and the $(-,+,+,+)$ metric signature. We also consider the value of the vacuum magnetic permeability $\mu_0$ to be equal to one.

\section{Theoretical setup}
\label{Sec:Theoretical}
\subsection{ Field equations}
We consider the model  of a complex scalar field $\Phi$, minimally coupled to Einstein’s gravity and coupled to Maxwell electrodynamics employing a generalization of the derivative operator. In this way, the action is given by 
\begin{equation}\label{eq:action}
S=\int d^4 x\sqrt{-g} \left[\frac{1}{16\pi }\mathcal{R}-\frac{\epsilon}{2}\left(g^{\mu\nu}(D_\mu\Phi)(D_\nu\Phi)^*+ \mu^2|\Phi|^2 - \frac{\lambda}{2}\,|\Phi|^4\right)-\frac{1}{4}F_{\mu\nu}F^{\mu\nu}\right],
\end{equation}
where $\mathcal{R}$ is the Ricci scalar, $\mu$ is the scalar field particle mass, $\lambda$ is the coupling constant, $F_{\mu\nu}=\partial_\mu A_\mu-\partial_\nu A_\mu$ is the Faraday tensor
and, as mentioned above, the covariant derivative operator, $D_\mu=\nabla_\mu+iqA_\mu$ couples the scalar field with the gauge field $A_\mu$ through the electromagnetic constant $q$. Here $\epsilon$ is equal to one when the scalar field is canonical and minus one when the scalar field describes phantom matter, and in this work, we will consider this last type of matter, so that $\epsilon=-1$. 
The scalar field defines the source current for the electromagnetic one, and in turn, the electromagnetic field also affects the geometry by means of Einstein's equations.
The Einstein-Klein-Gordon-Maxwell equations are obtained by taking a variation of Eq.~(\ref{eq:action}),
with respect to the different fields leading  to the Euler-Lagrange equations
of the model (see \textit{e.g.} \cite{Hawking:1973uf}).
The variation with respect to $g_{\mu\nu}$ leads to Einstein's equations:
\begin{subequations}\label{eq:ekgm}
\begin{eqnarray}
&& R_{\mu\nu}-\frac{1}{2}\mathcal{R} g_{\mu\nu}=8\pi T_{\mu\nu} \ ,
\label{eq:einstein} \\
&& T_{\mu\nu}={T^{\Phi}}_{\mu\nu}+{T^{\mathrm{EM}}}_{\mu\nu} \ , \label{eq:stress_energy}
\end{eqnarray}
\end{subequations}
where the stress energy tensors is given by
\begin{eqnarray}
&& {T^{\Phi}}_{\mu\nu}:=-\frac{1}{2}\,\left[(D
_{\mu}\Phi)(D_{\nu}\Phi)^*+ (D_{\nu}\Phi)(D_{\mu}\Phi)^*
-g_{\mu\nu}\left(g^{\alpha\beta}(D_{\alpha}\Phi)(D_{\beta}\Phi)^*+\mu^2|\Phi|^2+\frac{\lambda}{2}|\Phi|^{4}\right)\right], \label{eq:tmunu_esc}\\
&& {T^{\mathrm{EM}}}_{\mu\nu}:=\left(F_{\mu\sigma} F_{\nu\lambda}g^{\sigma\lambda}-\frac{1}{4} g_{\mu\nu} F_{\alpha\beta} F^{\alpha\beta}\right)\ . \label{eq:tmunu_em}
\end{eqnarray}

The variation of the action with respect to the scalar field $\Phi$ gives the Klein-Gordon equation,
\begin{equation}\label{eq:kg}
    g^{\mu\nu}D_\nu D_\mu \Phi=\mu^2\Phi - \lambda|\Phi|^2\Phi\ .
\end{equation}

Finally, the variation with respect to the electromagnetic potential $A_\mu$ leads to the Maxwell equations with the charged scalar field that defines the current four-vector and acts as a source of the electromagnetic field,
  \begin{eqnarray} \label{eq:maxwell}
    \nabla_\nu F^{\mu\nu}=J^\mu:=q j^\mu\ ,
  \end{eqnarray}
where the current is given by 
\begin{eqnarray}\label{eq:current}
  j^\mu=-\frac{i g^{\mu\nu}}{2}\left[{\Phi}^*(D_\nu\Phi)-\Phi(D_\nu\Phi)^*\right]\ ,
  \end{eqnarray}%
here $j^\mu$ is the Noether current of the complex field $\Phi$.

\subsubsection{Static spherically symetric spacetime and Ans\"atze for the fields}

We consider a line element with spherical symmetry in isotropic coordinates
\begin{equation}
    ds^2=-N^2 dt^2 + \Psi^4 \left[d\eta^2 + (\eta^2+{\eta_0}^2)\,d\Omega^2\right]\ , \label{eq:eleS}
\end{equation}
where the metric elements $N$ and $\Psi$ depend only on the radial coordinate   $\eta$, $d\Omega^{2}$ is the solid angle element and we have included a constant radius $\eta_0$ to have the wormhole feature of a non zero minimal radius.

In order to have no time dependence in Einstein equations, we assume for the complex scalar field the harmonic ansatz
\begin{equation} \label{eq:ansatz_phi}
    \Phi(\eta,t)= \phi(\eta) e^{i\omega t}\ ,
\end{equation}
where $\omega$ is a real constant.

Finally, consistent with the spherical symmetry, we consider that the gauge potential $A_\mu$ only has a temporal component, which is given by  

\begin{equation}
    A_\mu dx^\mu = V\left(\eta\right)\, dt\ , \label{eq:ansatz_V}
\end{equation}
where $V\left(\eta\right)$ defines the electric potential.
\subsubsection{ 3+1 decomposition of the stress energy tensor and system of field equations}

The Einstein equations can be written as a set of elliptic equations such that the sources are expressed  in terms of quantities like the energy density $\tau$, the momentum density $P_{\mu}$ and the stress tensor $S_{\mu\nu}$ explicitly. Let's start  with the 3+1 decomposition of the energy-momentum tensor, which consists of  the following projections of the stress energy tensor $T_{\mu\nu}$:
\begin{equation}
\tau=T_{\mu\nu}\,n^\mu\,n^\nu\ , \  \  P_\mu=-n^\sigma\,T_{\sigma\alpha}\, {\gamma^\alpha}_\mu \ , \  \ S_{\mu\nu}=T_{\alpha\beta}\,{\gamma^\alpha}_\mu\,{\gamma^\beta}_\nu \ , \label{eq:projections}
\end{equation} 
where ${\gamma^\mu}_\nu={\delta^\mu}_\nu + n^\mu\,n_\nu$ is the projection operator  and $n=\,\left(1/N , 0 , 0 , 0 \right)$ is the normal vector to the hypersurfaces.
Using the expressions for the tensors $T_{\mu \nu}$ Eq.~(\ref{eq:tmunu_esc}) and Eq.~(\ref{eq:tmunu_em}), into Eq.~(\ref{eq:projections})
we write down the above projected quantities explicitly: 
\begin{eqnarray}
    \tau&=& \frac{1}{2\Psi^4} \left( \frac{1}{N^2} V'^2-\phi'^2  \right) -\frac{\phi^2}{2} \left(\mu^2-\frac{\lambda\phi^2}{2} + \frac{(Vq+\omega)^2}{N^2} \right)\ , \\
    S^\eta_{\ \eta}&=& \frac{1}{2 \Psi^4} \left(-\frac{1}{N^2} V'^2- \phi'^2  \right)  +\frac{\phi^2}{2}  \left(\mu^2-\frac{\lambda\phi^2}{2} - \frac{(Vq+\omega)^2}{N^2} \right)\ , \\
    S^\theta_{\ \theta}&=&S^\phi_{\phi} =\frac{1}{2 \Psi^4} \left(-\frac{1}{N^2} V'^2
    + \phi'^2  \right)  +\frac{\phi^2}{2}  \left(\mu^2-\frac{\lambda\phi^2}{2} - \frac{(Vq+\omega)^2}{N^2} \right)\ ,
\end{eqnarray} 
and $P_\alpha$ is identically equal to zero in this case. Here and from now on we use the shorthand 
notation $f':= \frac{d f}{d \eta}$.

The  Einstein-Klein-Gordon-Maxwell system of equations Eq.~(\ref{eq:ekgm}) are thus given by a system of four elliptic equations,  two for the metric coefficients, one for the scalar field $\phi$ and one for the gauge potential $V$. A differential equation for the lapse function $N$ only, can be obtained from the combinations of Einstein equations $-{}^t_{\ t}+{}^\eta_{\ \eta}+{}^{\theta}_{\ \theta}+{}^{\varphi}_{\varphi}$ and the equation for the conformal factor $\Psi$ only is given by the ${}^t_{\ t}$ component of the Einstein's equations.

The system has the explicit form:
\begin{eqnarray} 
    &&\Delta_3\Psi + \frac{1}{4}\frac{\eta_0^2}{(\eta^2+\eta_0^2)^2} \Psi=-2\pi\,\Psi^5\,\tau\ , \label{eq:set_electrowormhole}
    \\
	&&\Delta_3 N+2\frac{\Psi'N'}{\Psi}  =4\pi\, N\,\Psi^4\left(\tau+S\right)\ , \label{eq:N_wh} \\
    &&\Delta_3\phi + 2\frac{\Psi' \phi'}{\Psi} + \frac{N' \phi'}{N} =\Psi^4\left(\mu^2-\lambda\,\phi^2-\left(\frac{qV+\omega}{N}\right)^2\right)\phi\ , \label{eq:set_elec_wh_phi}\\
	 &&\Delta_3 V + 2\frac{\Psi' V'}{\Psi} - \frac{N' V'}{N}
    =-\,q\,\Psi^4(q V+\omega)\phi^2 \ , \label{eq:set_elec_wh_V}
\end{eqnarray}
where we have used the operator definition $\Delta_3:= \frac{d^2}{d \eta^2} + \frac{2\eta}{\eta^2+\eta_0^2}  \frac{d }{d \eta}$.

Furthermore, we write down explicitly  the ${{}_{\eta}}^{\eta}$ component of the Einstein equations as an additional equation that will be necessary  as a constraint to solve numerically the set of equations
\begin{equation}
\frac{\Psi'^2}{\Psi^{2}}+\frac{N'\Psi'}{N\Psi}+\frac{\eta}{\eta^{2}+\eta_{0}^{2}}\left(\frac{\Psi'}{\Psi}+\frac{N'}{2N}\right)-\frac{\eta_{0}^{2}}{4\left(\eta^{2}+\eta_{0}^{2}\right)^{2}}=2\pi S_{\,\,\eta}^{\eta}\label{eq:const2} \ .
\end{equation}

And we have defined $S$ as the trace of the stress tensor, given by $S:= \gamma^{i j} S_{i j}={S^\eta}_\eta+{S^\theta}_\theta+{S^\varphi}_\varphi$, with the following explicit expression for $S$ and for the term appearing as sources in the equation for the lapse: 
\begin{eqnarray}
    S&=& \frac{1}{2 \Psi^4} \left(\frac{1}{N^2} V'^2+
    \phi'^2  \right) +\frac{3\phi^2}{2}  \left(\mu^2-\frac{\lambda\phi^2}{2} - \frac{(Vq+\omega)^2}{N^2} \right)\ , \\
    \tau + S &=& \frac{1}{\Psi^4 N^2} V'^2  + \phi^2 \left( \mu^2 - \frac{\lambda \phi^2}{2} -  \frac{2(Vq + \omega)^2}{N^2} \right)\  .
\end{eqnarray}

\subsection{Global quantities}

For a stationary and asymptotically flat spacetime, Komar expressions allow us to calculate global quantities \cite{Wald:1984rg}. In particular,  the total mass of a given spacetime can be computed using the following Komar expression: 
\begin{equation}\label{eq:KomarM}
  M_{\rm K}=\frac{1}{4\pi}\int_{\Sigma_t}R_{\mu\nu}n^\mu\xi^\nu dV \ ,
\end{equation}
where $\Sigma_t$  denotes a spacelike hypersurface, $n^{\mu}$ is the timelike vector normal to $\Sigma_t$ with  $n_{\mu}n^{\mu}=-1$, so that $n=\left(\frac{1}{N}, \vec{0}\right)$, $\xi=\partial_{t}=\left(1, \vec{0}\right)$ is the timelike Killing vector, $dV= \sqrt{\gamma}\, d\eta\, d\theta \,d\varphi$ is the volume element where $\gamma$ is the determinant of the spatial metric. 
In our case, $\xi^{\mu}=N n^{\mu}$ and using the Einstein equations, this expression can be rewritten as:
\begin{equation}\label{eq:M_Komar}
    M_{K} = \int \left(2 T_{\mu\nu}- T_{\,\,\alpha}^{\alpha} g_{\mu\nu} \right) n^{\mu}\xi^{\nu} \sqrt{\gamma}\, d\eta\,  d\theta \,d\varphi \ ,
\end{equation}
further, we have that $\left(2\, T_{\mu\nu}- T_{\,\alpha}^{\alpha}\ g_{\mu\nu} \right) n^{\mu}\xi^{\nu}=N\left(\, T_{\,\,\mu}^{\mu}-2T_{\,\,t}^t \right)$.

On the other hand, the total number of particles $\mathcal{N}$ can be obtained from the four-current $j_{\mu}$ defined in Eq.~(\ref{eq:current}). The current $j_{\mu}$ arises from the invariance of the action Eq.~(\ref{eq:action}) 
 under the global $U(1)$ transformations $\Phi\rightarrow\Phi e^{i\alpha},$ this implies that the current~(\ref{eq:current}) is Noether density current  and satisfies the conservation law $\nabla_{\mu}j^{\mu}=0$. Integration of the conserved law over a spacelike hypersurface $\Sigma_t$ defines the conserved Noether charge
\begin{equation}\label{eq:KomarN}
  \mathcal{N}=\int_{\Sigma_t} j^\mu n_\mu dV \ ,
\end{equation}
which can be associated with the total number of  particles \cite{Ruffini.187.1767}. The charge of the configuration can thus be defined as $Q=q \mathcal{N}$. On the other hand, using the Komar mass and the particle number expressed in terms  of the volume integral,  Eqs.~(\ref{eq:KomarM}) and Eq.~(\ref{eq:KomarN}), it is possible to compute the  $M_{k}$ from the gradient of the lapse function $N$ on a 2-sphere at spatial infinity (see \textit{e.g.} \cite{Gourgoulhon:2010}), while the charge $Q$ can be computed from the gradient of the gauge potential $V$. Therefore, the global quantities $M_{k}$  and $Q$ can be extracted from the  asymptotic behavior of the metric \eqref{eq:eleS} and of the gauge potential $V(\eta)$, as $M_{\rm K}= \lim_{\eta\to\infty}\eta^2 N'$ and  $Q=4\pi\lim_{\eta\to\infty}\eta^2 V'$.

\section{Solutions }
\label{Sec:Solutions}

\subsection{Boundary conditions }

In order to construct the electrostatic solution that describes a wormhole it is necessary to  fix the parameters $\{\omega,\lambda, q\}$, and solve the  system of differential equations for the functions  $\{ N ,\Psi , \phi , V \}$  by imposing appropriate  boundary conditions on the scalar field, the gauge potential and the metric functions. 
We impose reflection symmetry on the throat, at $\eta=0$, so that the functions must satisfy the following: 
\begin{equation}\label{eq:regularity_eta}
   N' |_{\eta=0}=0, \quad \Psi' |_{\eta=0}=0, \quad \phi'|_{\eta=0}=0, \quad V'|_{\eta= 0}=0
     \ ,
\end{equation}
and demanding asymptotic flatness implies
\begin{equation}\label{eq:out_bc}
    N|_{\eta\to\infty}=1, \quad \Psi|_{\eta\to\infty}=1, \quad \phi|_{\eta\to\infty}=0,\quad V|_{\eta\to\infty}=0 \ , 
\end{equation}
the asymptotic vanishing of the scalar field implies the condition $\omega^2 < \mu^2$.

Additionally, these  boundary conditions imply a constraint among the functions evaluated at the throat and the system parameters. To see this, we use the circumferential radius $R$ which is given by
\begin{equation}
R= \Psi^2 \sqrt{\eta^2+{\eta_0}^2}\ ,  
\end{equation}
where the minimal circumference, {\it i.e.}, the throat of the wormhole is at $\eta=0$, so that the radius of the throat is given by ${\cal G}=\Psi_c^2 \eta_0$.
 On the other hand, another expression for such throat radius can be obtained from the ${}_{\ \eta}^{\eta}$ component of the Einstein equations, Eq.~(\ref{eq:const2}).  Indeed, evaluating such expression at the throat, $\eta=0$, and using the boundary conditions, we obtain another expression for the throat radius and equating it with the previous one, we derive a constraint equation that allows us to determine the parameter $\eta_0$ in terms of the other parameters and the values of the functions at the throat:

\begin{equation} \label{eq:constriction}
    {\cal G}^2=\Psi_c^4 \eta_0^2 = \frac{1}{4\pi \phi_c^2\left(-\mu^2+\frac{\lambda\phi_c^2}{2} + \frac{\left(V_c\,q+\omega \right)^2}{N_c^2}\right)}\ ,
\end{equation}
where $N_c:=N(0)>0$, $\Psi_c:=\Psi(0)$, $\phi_c:=\phi(0)>0$, and $V_c:=V(0)$ are the central values for the metric coefficients $N$, $\Psi$, the scalar field $\phi$ and the electric potential $V$ respectively. It is a constraint equation among the parameters using the boundary conditions of the system. 

\subsection{Numerical setup and particular solutions}
The non-linear system of PDEs (\ref{eq:set_electrowormhole}-\ref{eq:set_elec_wh_V}) is solved numerically together  with boundary conditions (\ref{eq:regularity_eta}) and (\ref{eq:out_bc}) using a spectral collocation method with Chebyshev polynomials as a spectral basis for the unknown functions  $\{ N ,\Psi , \phi , V \}$  in a compactified domain. For details on the method, see for instance Ref.~\cite{Grandclement:2007sb}. The solutions presented in this work have been found by means of a Newton-Raphson iteration.

We have constructed numerically several wormhole solutions, varying the parameter $\lambda$ in the interval $\frac{1}2 \leq \lambda/4\pi \leq  100$, the boson frequency $\omega$ is explored in the interval $0\leq \omega/\mu \leq 1$, and we analyze cases for a fixed value of $q$, increasing it  gradually between $0\leq q/\sqrt{8\pi} \leq  0.5$. Such procedure continues the one given at \cite{Dzhunushaliev:2008bq}, and these intervals are chosen in order to obtain regular physically acceptable solutions.

Now, let us start discussing the general behaviour of the electric wormhole. As mentioned above, we obtain numerical solutions for different values of the parameters $\omega$, $\lambda$ and $q$. Using the invariance of the equations (\ref{eq:set_electrowormhole}-\ref{eq:set_elec_wh_V}) under the scaling:
\begin{equation}\label{eq:tilde}
    r \rightarrow   \mu r , \quad \omega \rightarrow   \omega/\mu, \quad 
    \lambda   \rightarrow  \lambda/\mu^2 , \quad    q   \rightarrow   q/\mu\ , 
\end{equation}
we thus obtained solutions for arbitrary values of $\mu$. All the further reported quantities will be given in terms of the mass of the scalar field $\mu$. 
Additionally, we use the following scaling for the charge and scalar field  to facilitate a comparison with the charge and scalar field scale reported in the literature, \textit{e.g.} \cite{Chew:2019lsa} 
\begin{equation}
   \tilde{q} =\frac{q}{ \sqrt{8\pi} },\quad\tilde{\phi}=\frac{\phi}{\sqrt{4 \pi}},\quad\tilde{\lambda}=\frac{\lambda}{4 \pi} \ . \label{eqs:factors}
\end{equation} 

\begin{figure}
	\begin{centering}
		\includegraphics[scale=.5]{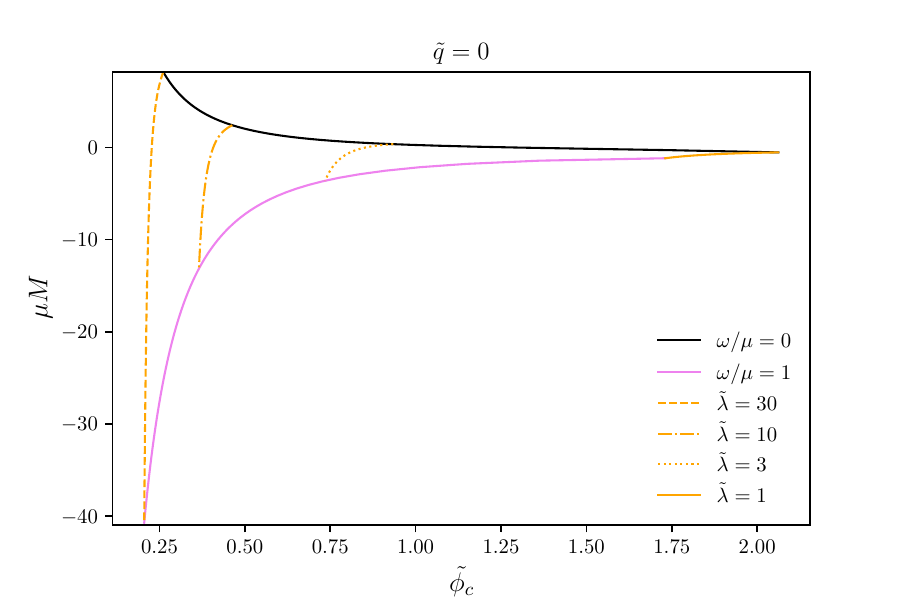}
            \includegraphics[scale=.5]{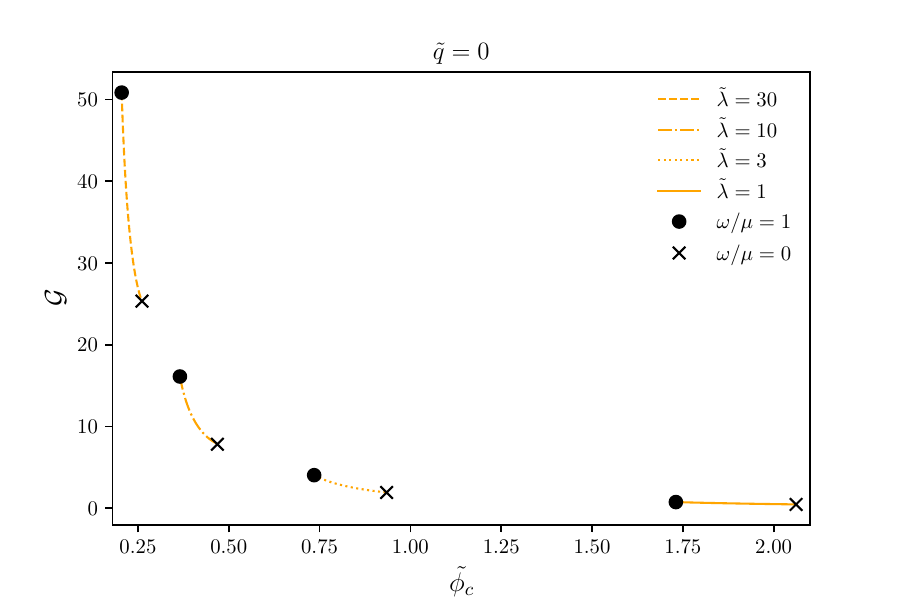}
            \includegraphics[scale=.5]{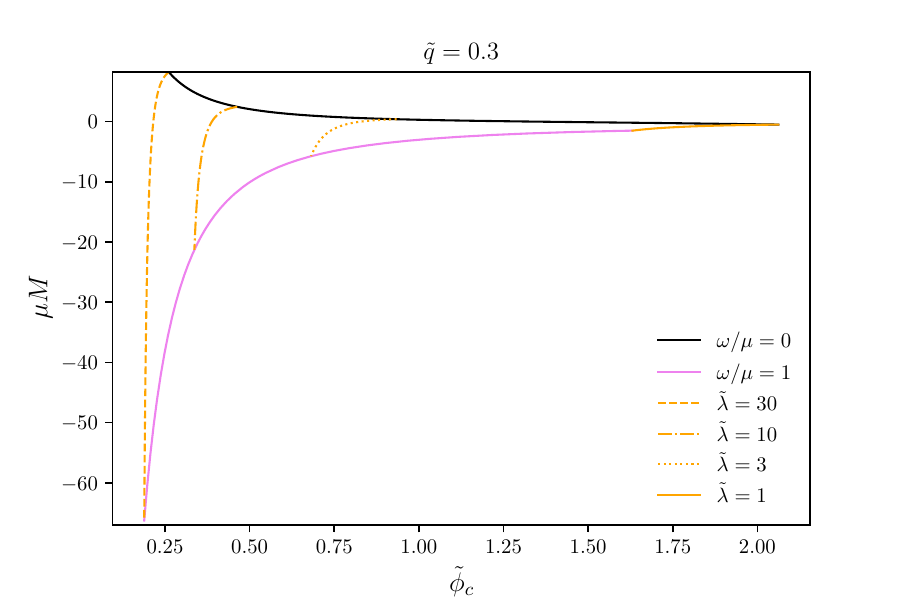}
            \includegraphics[scale=.5]{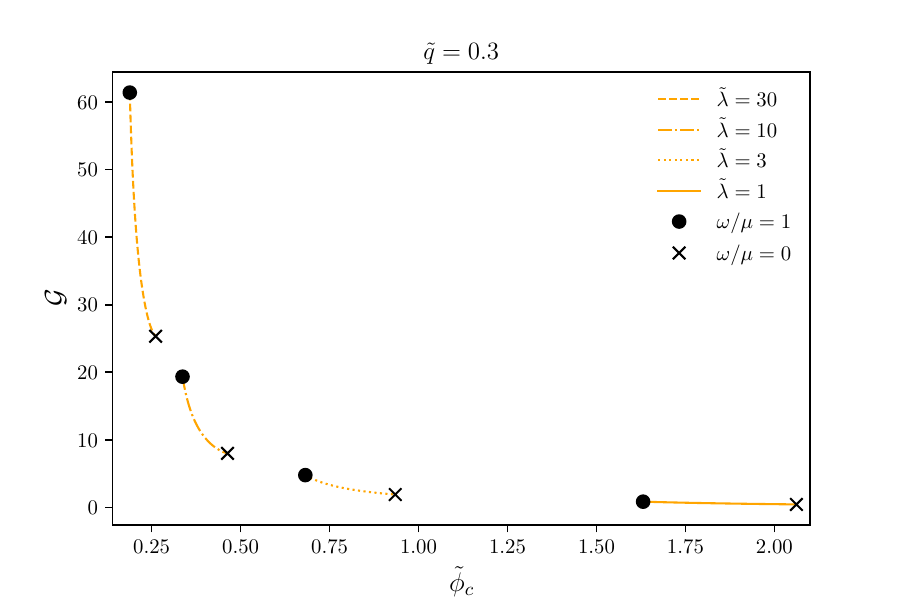}
            \includegraphics[scale=.5]{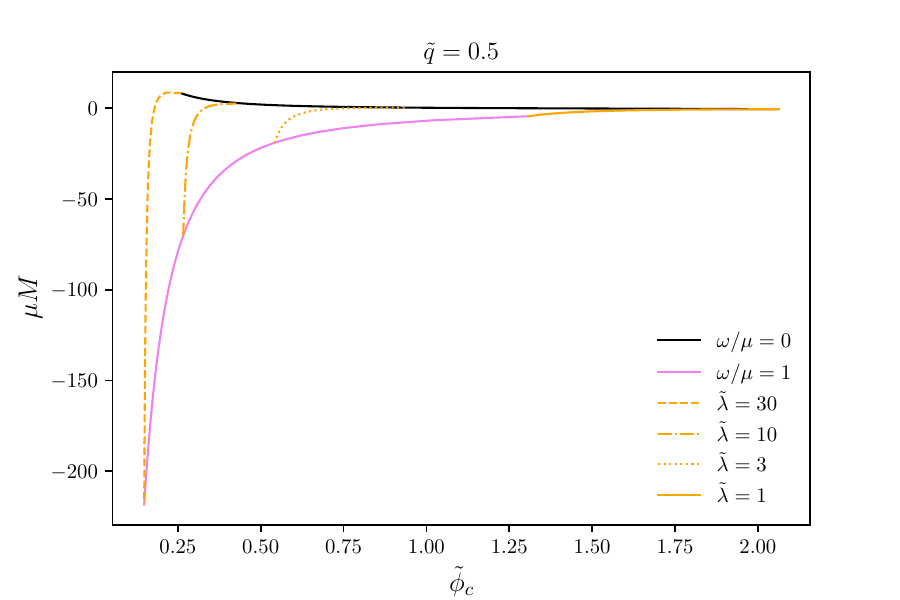}
            \includegraphics[scale=.5]{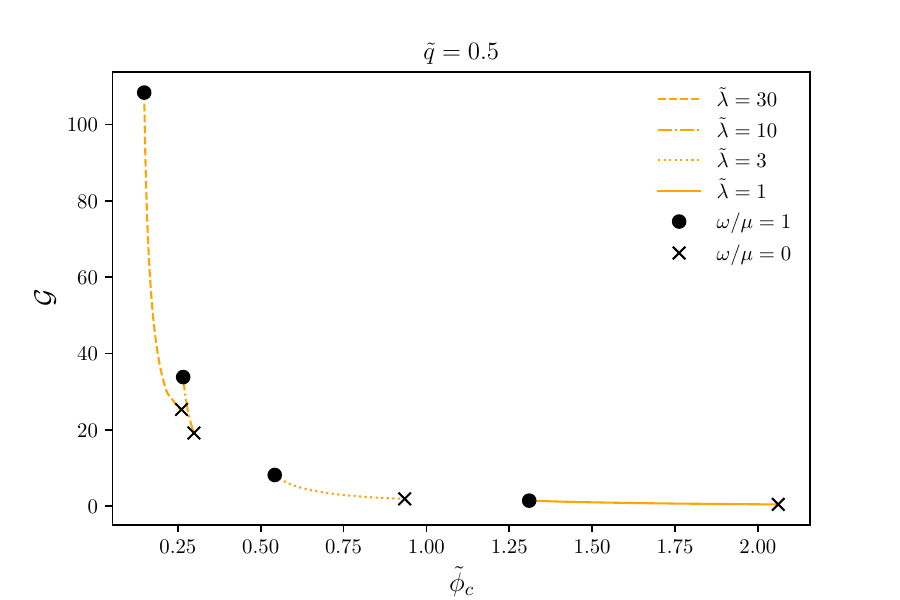}
		\par
	\end{centering}
	\caption{In the left panels, we plot the wormhole total mass $M$ and in the right one the wormhole radius as a function of the central value of the scalar field $\tilde{\phi}_c$. Notice that $\tilde{\lambda}$ and $\tilde{\phi}_c$ are monotonically related, for each value of $\phi_c$ corresponds to one of $\tilde{\lambda}$.
 The solid lines are solutions for $\omega/\mu=0$ (black) and $\omega/\mu=1$ (violet). We have also included cases for different fixed values of the parameter $\tilde{\lambda}$ (orange) where the frequency takes values between $0$ and $1$, with $\tilde{\lambda} \in[ 1,30]$. }
\label{fig:phic_M_T}
\end{figure}

In Fig.~\ref{fig:phic_M_T}, we present  solutions for the total mass $M$ and the throat radius, ${\cal G}$ of the electric wormhole  as a function of the central value of the scalar field $\phi_c$. Furthermore,   
we have used the values of the boson frequency $\omega/\mu=0$ (black lines) and $\omega/\mu=1$ (violet lines) varying $\tilde{\lambda}\in[1, 30]$. We have also included orange dashed lines for representative values of the $\tilde{\lambda}$-parameter varying $\omega/\mu\in[0,1]$.
 
The $\omega/\mu=0$ curve delimits the maximum mass of wormholes regardless of the value of $q$, while the $\omega/{\mu}=1$ curve does vary with $q$. When $\omega/\mu \sim 1$, as $q$ increases, the total mass and radius of the throat also increase in magnitude when $q$ increases. 

The  solutions with ${\tilde{q}}=0$ are consistent with  the results in \cite{Dzhunushaliev:2008bq} as can be seen by comparing their Figs. 2 and 3 with our first row in Fig.~\ref{fig:phic_M_T}. Solutions with $\tilde{q}>0$ follow the same qualitative relations between the global quantities that the non-charged wormholes. We also present how the wormhole mass and throat radius depend  on the central value of the scalar field. The difference with the neutral case is that the charged wormholes with close to one boson frequencies $\omega/\mu$  reach bigger (negative) masses and larger throat radii than the corresponding  non-charged wormholes.  The common behavior for the values of the charged analyzed is that for small central values of the scalar field, $\tilde{\phi}_c$, the throat radius increases almost exponentially whereas, for large central values, the throat radius tends to zero. For the neutral cases, the total mass for $\omega/\mu=0$ is always positive, whereas for $\omega/\mu=1$ is always negative, and this behavior changes with the charge, see below. Also notice that the throat radius for a given value of the scalar field at the throat is always larger for $\omega/\mu=1$ than the corresponding value for $\omega/\mu=0$, a feature that remains in the charged cases. We have also included cases for different values of the parameter $\tilde{\lambda}$, where the frequencies vary from $\omega/\mu=0$ to $\omega/\mu =1$. 

\begin{figure}
	\begin{centering}
		\includegraphics[scale=.5]{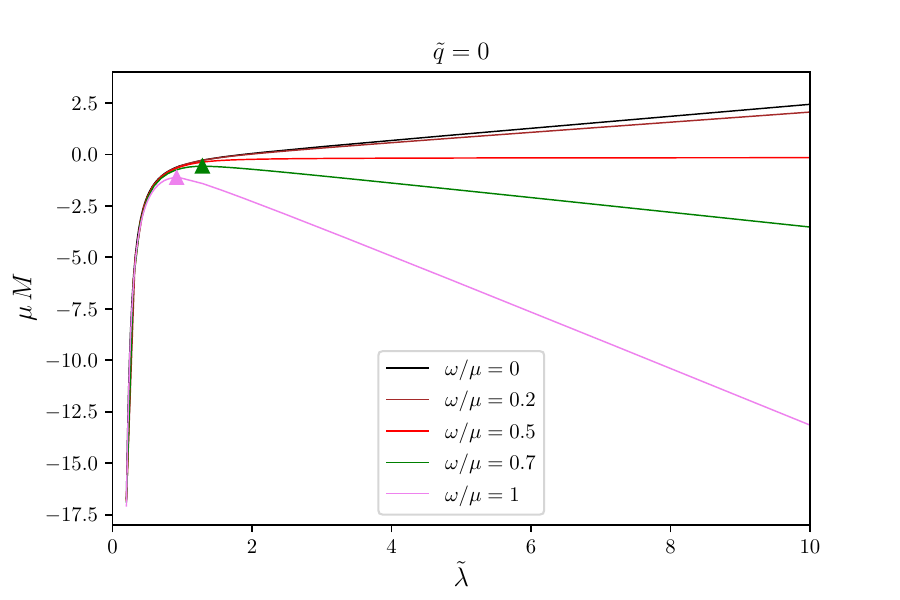}
            \includegraphics[scale=.5]{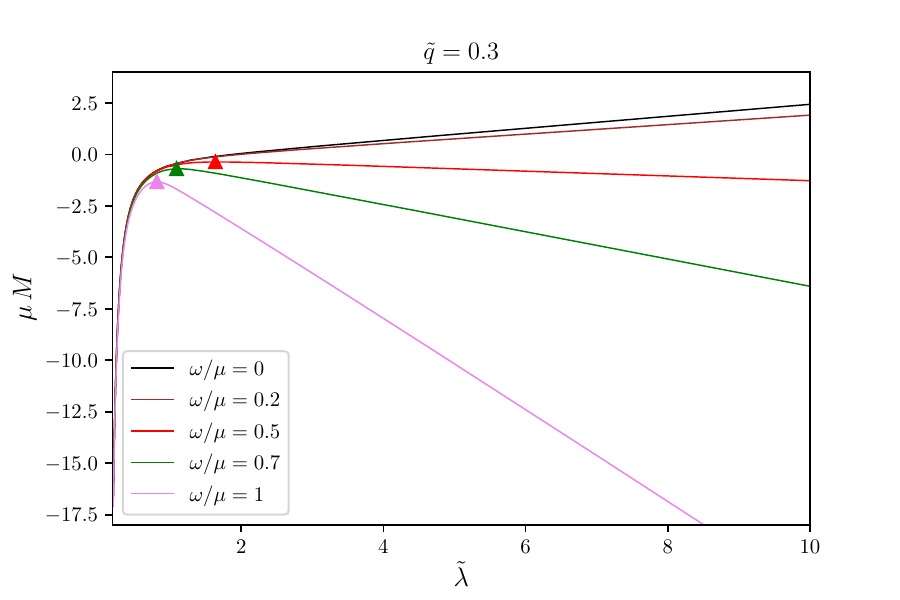}
            \includegraphics[scale=.5]{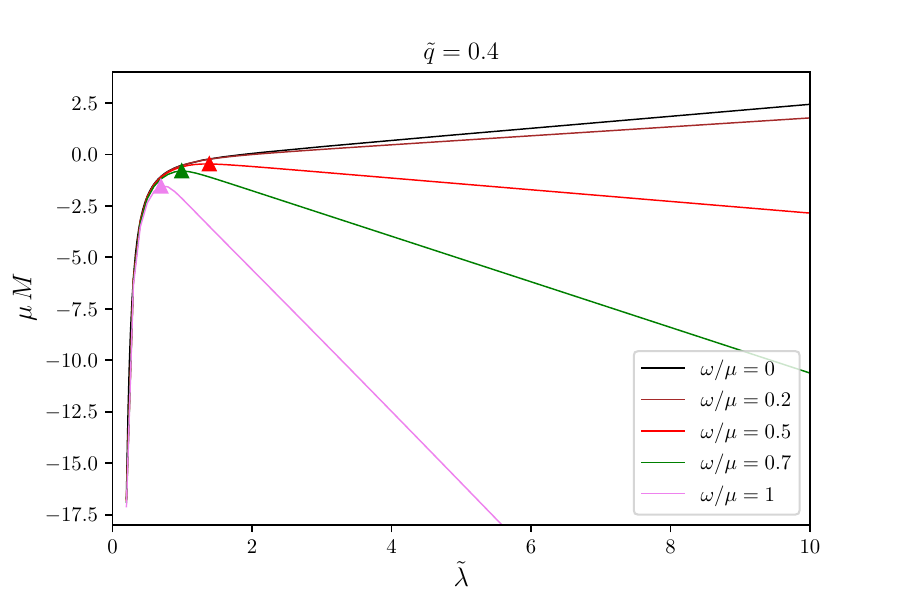}
             \includegraphics[scale=.5]{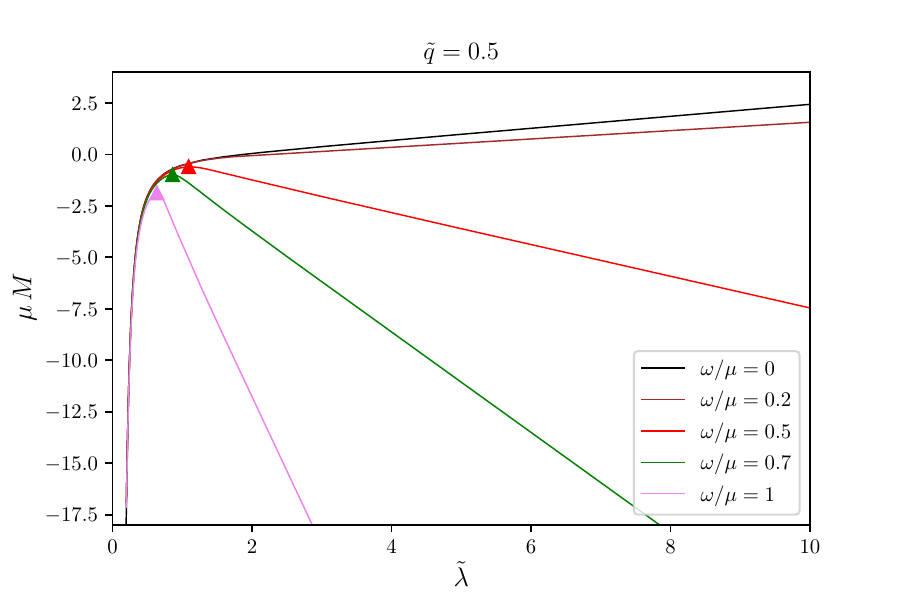}
		\par
	\end{centering}
	\caption{ 
        The mass of wormhole solutions versus $\tilde{\lambda}$. We show the mass $M$ as a function of $\tilde{\lambda} >0$ in the full interval of the boson frequency $\omega/\mu\in[0,1]$ for different values of the electric charge of the scalar field  $\tilde{q}\in \left[ 0,\, 0.5\right]$. The inverted triangles represent the maximum of the mass $M_{max}$ of the electric wormhole for different  frequencies. The values of the parameters are given in Table~\ref{tab:Tab1}.  } 
	\label{fig:Mvslambda}
\end{figure}

\begin{figure}
	\begin{centering}	
            \includegraphics[scale=.5]{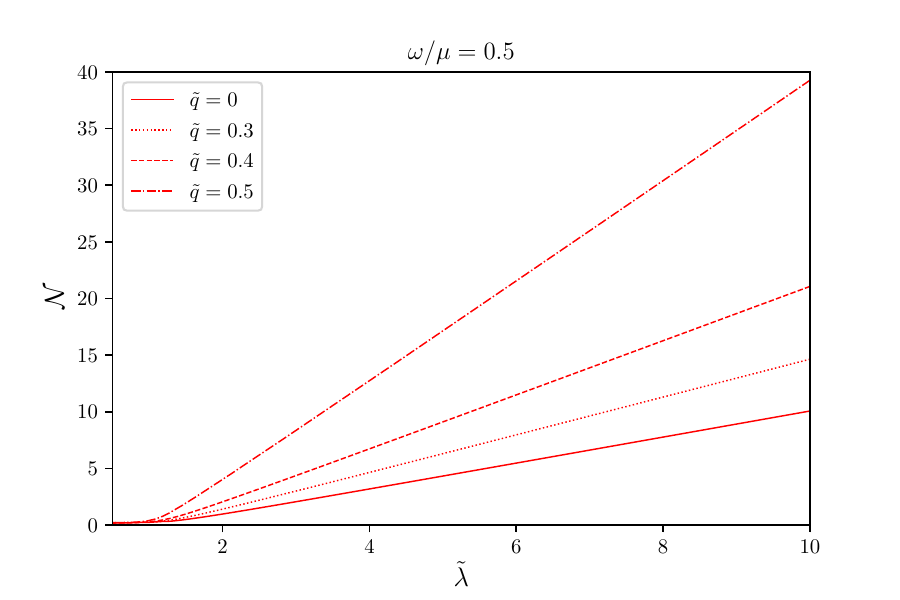}
            \includegraphics[scale=.5]{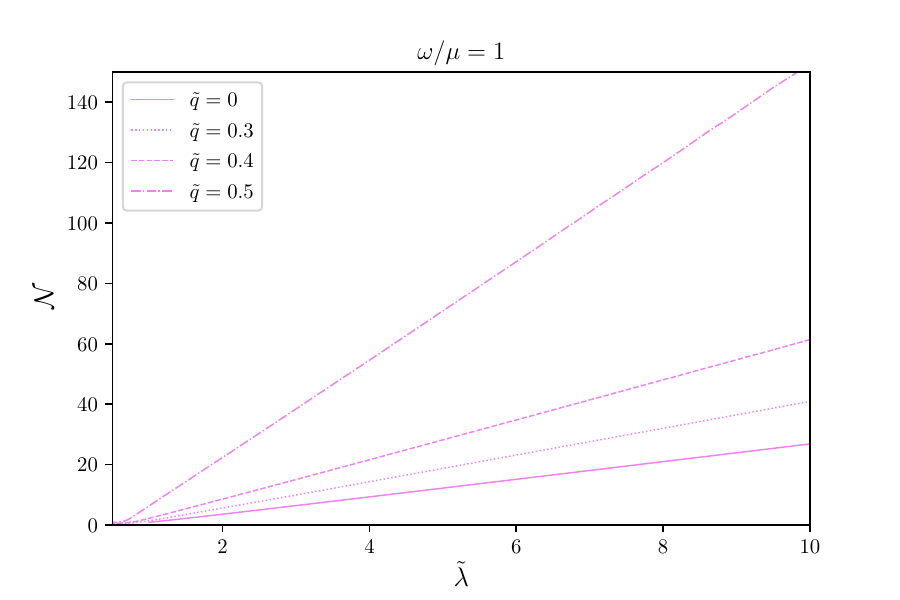}
            
		\par
	\end{centering}
	\caption{ Number of particles of the charged wormhole as a function of the parameter $\tilde{\lambda}$  for some representative values of $ \tilde{q}$. The solid line represents the non-charged case and the dotted lines represent the charged solutions.} 
	\label{fig:N_lambda}
\end{figure}

In order to highlight the behaviour of the charge $q$ on the configurations, in the code, we fixed a value of the charge $q$, analysed cases for particular values of the parameter $\lambda$ and the frequency $\omega$, and repeated the procedure for another value of the charge. In \cite{Chew:2019lsa}, the authors analyzed the properties of the non-charged wormhole solutions where their solutions were focused on the behavior of $\lambda$. Following this procedure, in Fig.~\ref{fig:Mvslambda},  we plot the total mass $M$ as a function of $\tilde{\lambda}$ for several values of $\omega/\mu$, and in Fig.~\ref{fig:N_lambda}, we show the number of particles  $\mathcal{N}$ as a function of $\tilde{\lambda}$ for several values of the charge, with a fixed value of $\omega/\mu$, to emphasize  the differences between the charged wormholes with the non-charged ones, which were discussed in \cite{Chew:2019lsa}. 


In Fig.~\ref{fig:Mvslambda}, we see that the behavior of the total mass as a function of $\tilde{\lambda}$  in the charged case, is similar to one of the neutral case: for small values of $\tilde{\lambda}$, is negative and essentially independent of the frequency $\omega/\mu$, and in this case, see Fig.~\ref{fig:phic_M_T}, the throat radius tends to zero and the wormhole is closing. As $\tilde{\lambda}$ increases, the $\omega/\mu=0$ wormhole acquires a positive total mass that increases lineally with $\tilde{\lambda}$. 

There is a frequency, near $\omega/\mu=0.5$ for which the total mass is equal to zero for any value of $\tilde{\lambda}$ greater than  $\sim 2$; as the frequency increases, the total mass reaches a maximum value and then becomes more negative, with a linear dependence on $\tilde{\lambda}$. 

The presence of the charge maintains the general behavior of the mass as a function of  $\tilde{\lambda}$ but with noticeable differences. The $\omega/\mu=0$ configuration is independent of the charge, which is an expected result, as long as this case corresponds to a real scalar field that has no possibility of being charged. For $\omega/\mu \neq 0$, the role of the charge is very noticeable, changing the value of the frequency at which the total mass acquires the zero value and becomes essentially independent of $\tilde{\lambda}$, and making the slope of the dependence of the total mass as a function of $\tilde{\lambda}$ much more pronounced. 

We have focused on the specific values of $\tilde{\lambda}$ at which the total mass reaches a maximum, for the several values of the charge and the different frequencies. In Fig.~\ref{fig:Mvslambda}, we present plots of the total mass $M$ for different values of the boson frequency $\omega/\mu$ given a value of the charge, and panels for the different charge values, for which we choose, $\tilde{q}=0,\ 0.3,$ and $0.5$.
We have marked the points of maximum mass on the different plots and written them in Table.~\ref{tab:Tab1}. Also, we have identified the frequency which yields the case mentioned above of a total mass equal to zero, for each value of the charge it is a particular frequency, and we label it the {\it zero mass frequency},  $\omega_{\rm \rm zm}$.

For solutions with $\omega >$ $\omega_{\rm \rm zm}$, there aren't local  maximum for the mass, the mass increase linearly with $\tilde{\lambda}$, whereas for $\omega<$ $\omega_{\rm \rm zm}$, the total mass  grows with $\lambda$, reaches a maximum value, and then linearly decreases.
In the case of $\tilde{q}=0$, the zero mass frequency  is $\omega_{\rm \rm zm}/\mu=0.5$, and the mass increases (decreases) linearly for $\omega/\mu>0.5$ ($\omega/\mu<0.5$) as was reported in \cite{Chew:2019lsa}. 

The values of the frequency  $\omega_{\rm \rm zm}$ decrease for the charged wormhole; those corresponding to configurations with $\tilde{q} = 0$, $0.3$, $0.4$, and $0.5$ are given in Table.~\ref{tab:Tab1},
where $\omega_{\rm \rm zm}$ corresponds to the  value of the frequency such that the mass becomes zero.
\begin{SCtable}
\begin{tabular}{cccccccc}
\hline 
$\tilde{q}$ &  & $\omega/\mu$ &  & $\mu M_{max}$ &  & $\tilde{\lambda}$ & \tabularnewline
\hline 
\hline 
$0$ &  & $0.5^*$ &  & $0$ &  & $\infty$ & \tabularnewline
$0$ &  & $0.7$ &  & $-0.578$ &  & $1.289$ & \tabularnewline
$0$ &  & $1$ &  & $-1.132$ &  & $0.919$ & \tabularnewline
$0.3$ &  & $0.4375^*$ &  & $0$ &  & $\infty$ & \tabularnewline
$0.3$ &  & $0.5$ &  & $-0.3678$ &  & $1.636$ & \tabularnewline
$0.3$ &  & $0.7$ &  & $-0.6978$ &  & $1.091$ & \tabularnewline
$0.3$ &  & $1$ &  & $-1.3305$ &  & $0.814$ & \tabularnewline
$0.4$ &  & $0.3875^*$ &  & $0$ &  & $\infty$ & \tabularnewline
$0.4$ &  & $0.5$ &  & $-0.467$ &  & $1.388$ & \tabularnewline
$0.4$ &  & $0.7$ &  & $-0.809$ &  & $0.992$ & \tabularnewline
$0.4$ &  & $1$ &  & $-1.554$ &  & $0.694$ & \tabularnewline
$0.5$ &  & $0.3125^*$ &  & $0$ &  & $\infty$ & \tabularnewline
$0.5$ &  & $0.5$ &  & $-0.604$ &  & $1.091$ & \tabularnewline
$0.5$ &  & $0.7$ &  & $-0.995$ &  & $0.861$ & \tabularnewline
$0.5$ &  & $1$ &  & $-1.877$ &  & $0.636$ & \tabularnewline
\hline 
\end{tabular}
\caption{\label{tab:Tab1} The local maximum of the mass $M_{\mathrm{max}}$ for a given value of the charge $q$ and frequency $\omega\leq \omega_{\rm \rm zm}$. Configurations with $M_{max}=0$ define the zero mass frequency $\omega_{\rm \rm zm}$ when the frequency $\omega>\omega_{\rm \rm zm}$ the mass doesn't have a local maximum and increases linearly with $\lambda$.
}
\end{SCtable}

Regarding the total number of particles, Eq.~(\ref{eq:KomarN}), and the effect of the charge of the scalar field, $q$, 
on that global quantity, in Fig.~\ref{fig:N_lambda} we describe, given a frequency, the dependence of the number of particles of the configuration on the parameter $\lambda$, for several values of the scalar field charge $q$. As in the case of the total mass, for values of $\tilde{\lambda}$ not small (greater than $\sim 1.5$), the number of particles $\mathcal{N}$ increases  almost linearly with $\tilde{\lambda}$,  for a fixed value of $q$. The effect of the charge $q$ is that the slope of the particle number $\mathcal{N}$ as a function of $\tilde{\lambda}$ increases with $\tilde{q}$, reaching a maximum slope for $\tilde{q}=0.5$, and solving the system of equations for values of the charge larger than this, becomes computationally very demanding, possibly indicating the fact that there is a maximal value for the charge $\tilde{q}$ beyond which there are no static spherical configurations.
We present the case for two values of the frequency.  Both cases have similar behavior on the number of particles as a function of $\tilde{\lambda}$, with increasing slope for larger values of the charge $\tilde{q}$, but the number of particles, given $\tilde{\lambda}$ and $q$ is much larger than the corresponding value in the neutral case, for the larger value of the boson frequency $\omega/\mu$.

We notice that with the expression that we are using, Eq.~(\ref{eq:KomarN}), we can not obtain the number of particles related to the negative mass and the corresponding one related to the positive one; it would be interesting to derive an expression that differentiates such numbers. For this expression, it would seem that the particles associated with the positive mass and those associated with the negative mass, are counted in the same way.

\begin{figure}[ht]
	\begin{centering}
             \includegraphics[scale=.47]{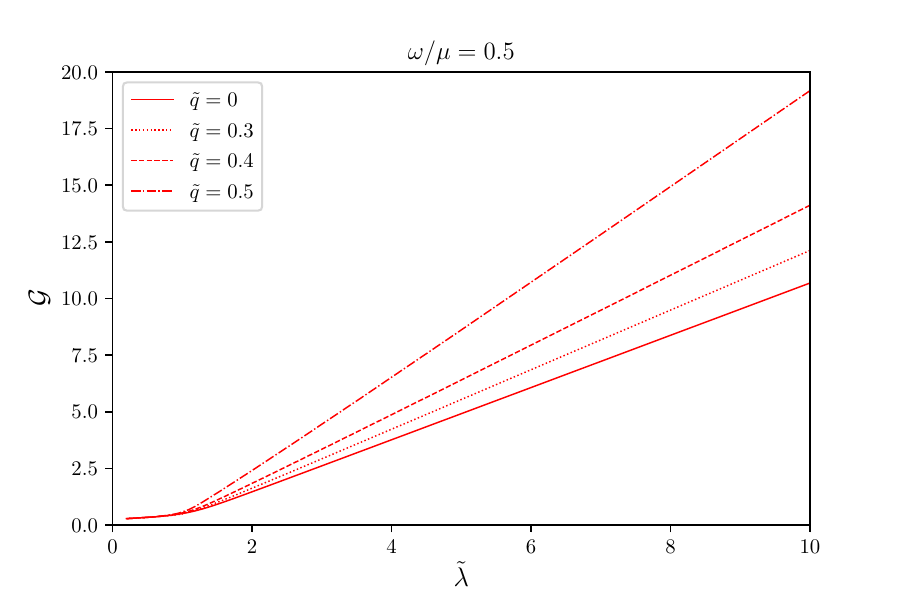}
            \includegraphics[scale=.47]{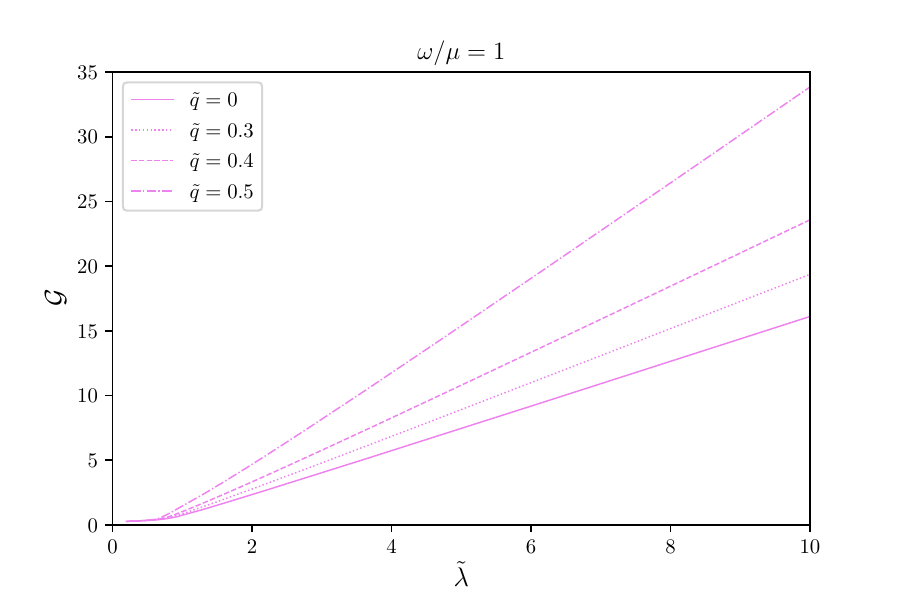} 
		\par
	\end{centering}
	\caption{ We present some numerical solutions for the throat ${\cal G}$ as a  function of $\tilde{\lambda}$ for different values of the charge $q$ and  two values of $\omega/\mu $. The size throat grows linearly with $\tilde{\lambda}$, when the parameter is not too small, and the slope increases with $q$. The actual size of the throat increases with $\omega$.
} \label{fig:eta0_lambda} 
\end{figure}

In Fig.~\ref{fig:eta0_lambda}, we present the throat radius
${\cal G}$ as a function of $\tilde{\lambda}$ and for several values of the scalar charge $\tilde{q}$ for two values of the boson frequency ${\omega}/\mu$. It is remarkable that the throat radius profile is very similar to the one of the number of particles, possibly indicating a relation between this feature of the wormhole and that global quantity. Another interesting feature of the charged wormhole was that it presented the possibility of challenging the theorem valid for non-charged wormholes 
\cite{Carvente:2019gkd}, stating that the presence of the self-interaction parameter, $\tilde{\lambda}$ is a necessary condition for having a wormhole. We explore the possibility that the charge, $\tilde{q}$, could play the role of the self-interaction parameter and obtain charged wormholes without it. Our numerical experiments showed that this is not the case and that the presence of the self-interaction parameter remains a necessary condition for the existence of a wormhole, even in the charged case.  

\begin{figure}[ht]
	\begin{centering}
		\includegraphics[scale=.49]{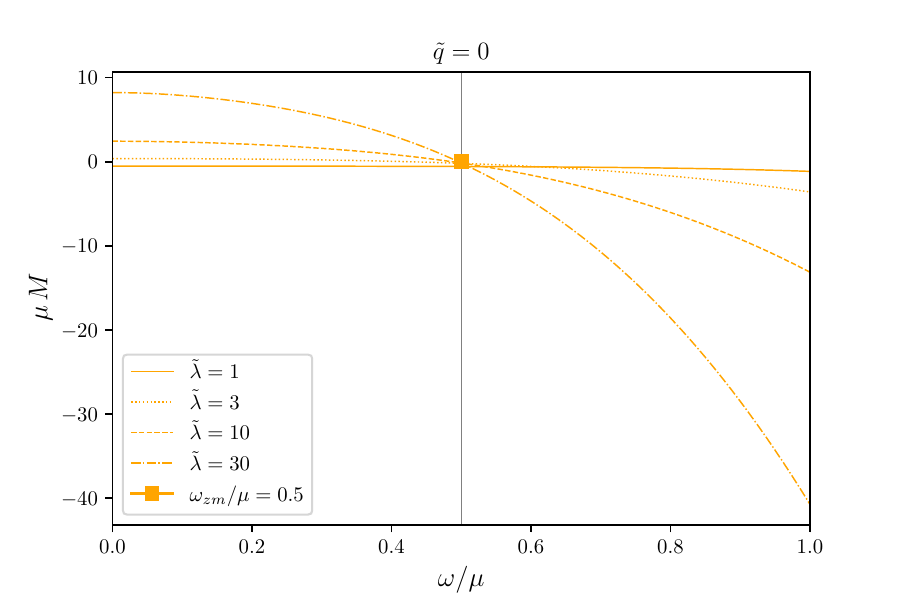}
            \includegraphics[scale=.49]{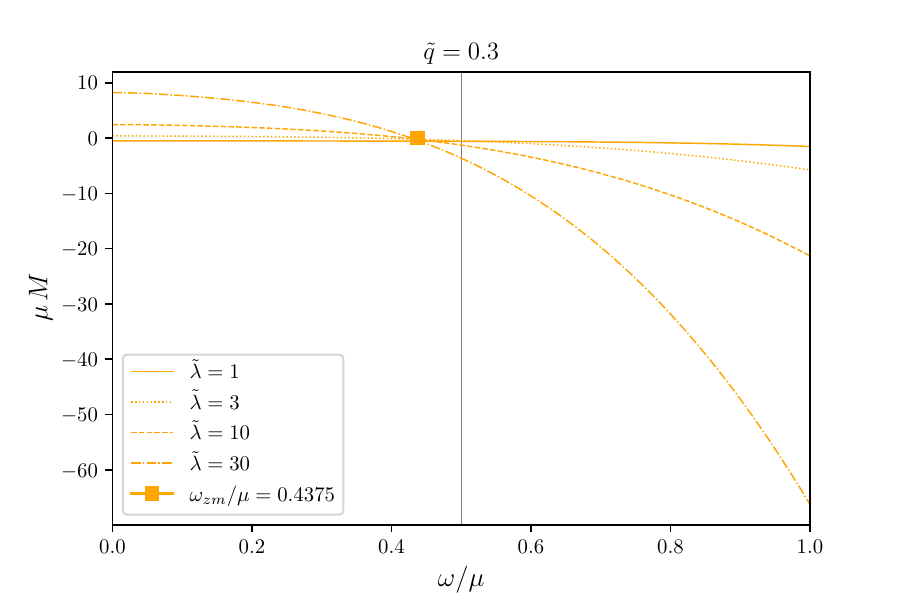}
            \includegraphics[scale=.49]{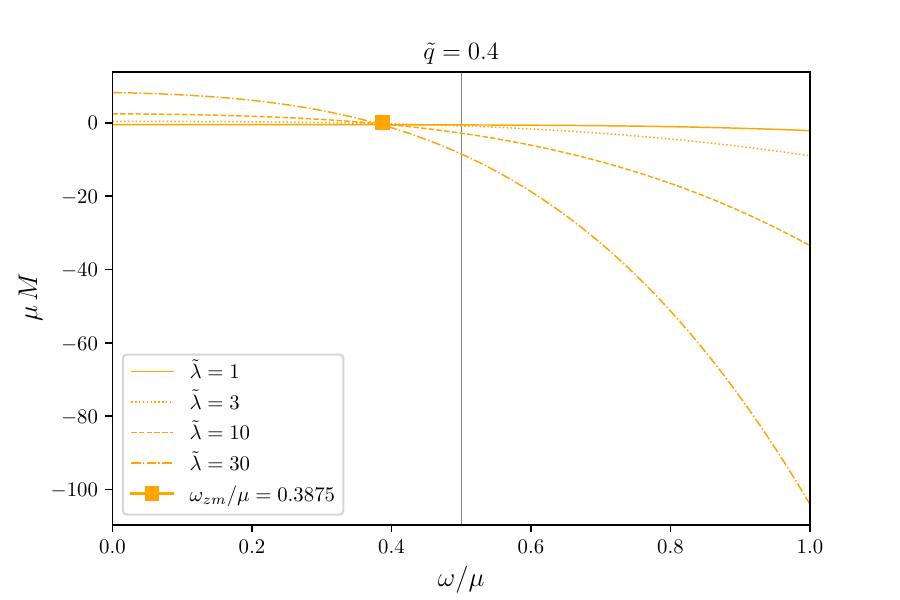}
             \includegraphics[scale=.49]{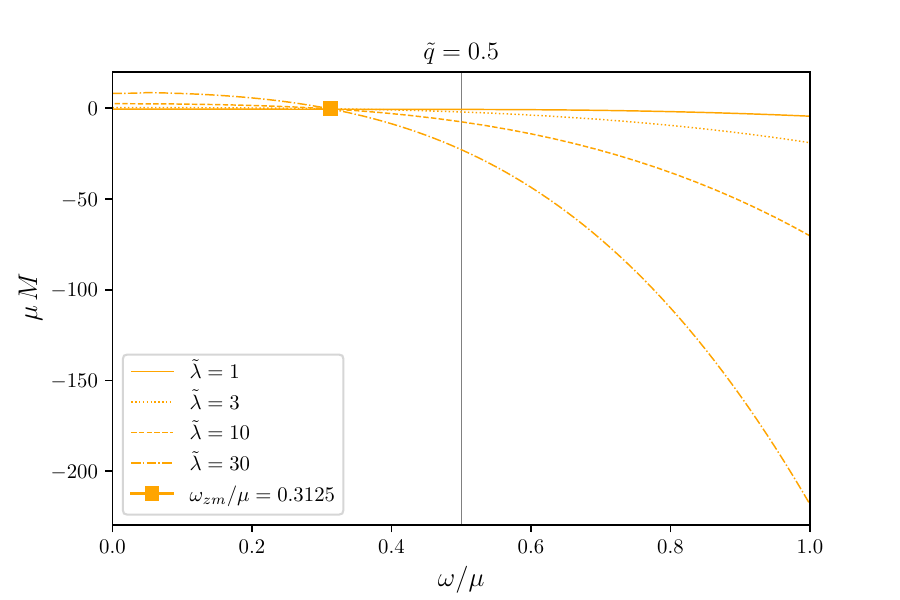}
		\par
	\end{centering}
	\caption{  The mass of a spherical electrical wormhole is shown for some values of the parameter $\tilde{\lambda}$ with different values of $\tilde{q}$.  Although the profiles are qualitatively similar, in the limit $\tilde{\lambda}\rightarrow\infty$ the mass becomes zero for a particular frequency defined as flat space frequency $\omega_{\rm \rm zm}/\mu$.  The square symbol corresponds  to such frequency for different charges $\tilde{q}$ and they are also reported in Table~\ref{tab:Tab1}. } 
	\label{fig:omegavsM}
\end{figure}

Finally, in Fig.~\ref{fig:omegavsM}, we present again the behavior of the total mass, now as a function of the boson frequency $\omega$ for several values of $\tilde{\lambda}$ and certain given values of the scalar charge, $q$.

With the factors given in Eq.~(\ref{eqs:factors}), we are able to compare with the results of previous works in the literature, as in \cite{Dzhunushaliev:2008bq} for the neutral case. As mentioned above, generically, one of the effects of the charge is to change the value of the boson frequency that gives a zero Komar mass; we have called such frequency {\it zero mass frequency}.

In the four panels of  Fig.~\ref{fig:omegavsM}, we have drawn a vertical line indicating the value $\omega_{\rm \rm zm}/\mu=0.5$, the value corresponding to the neutral case, and it is clearly seen how the effect of the charge is to reduce the value of the boson frequency which gives a zero Komar mass; furthermore, notice that it is a value independent of $\tilde{\lambda}$ for a given charge. Noticing that the system of field equations, Eqs.~(\ref{eq:set_electrowormhole}), is invariant under  the simultaneous change in signs of the charge $\tilde{q}$ and the electric potential, $V$, we see that there is no change in the total mass behavior due to a change of the sign of the scalar charge $\tilde{q}$. 

\begin{figure}[ht]
	\begin{centering}
		\includegraphics[scale=.5]{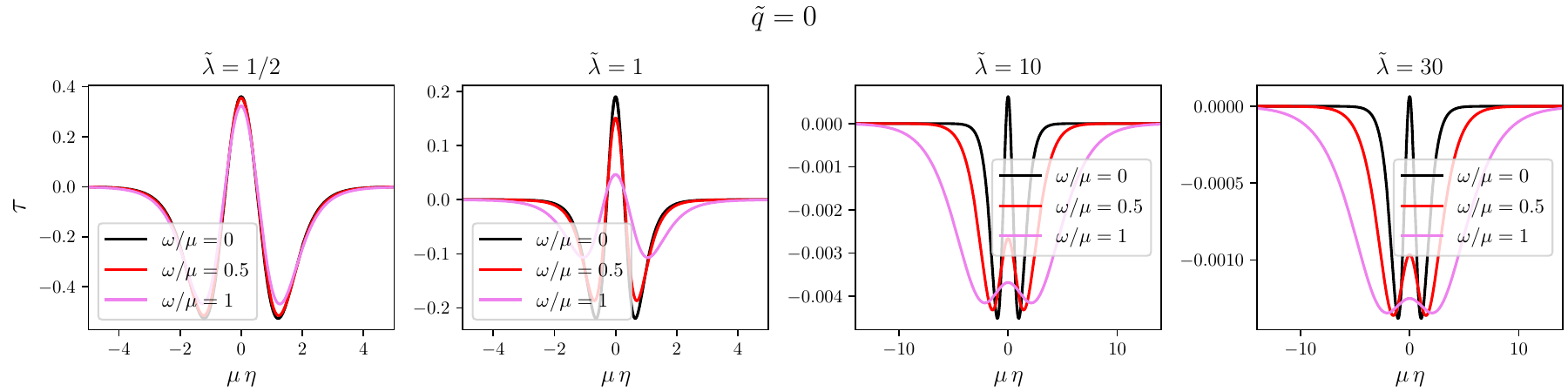}
            \includegraphics[scale=.5]{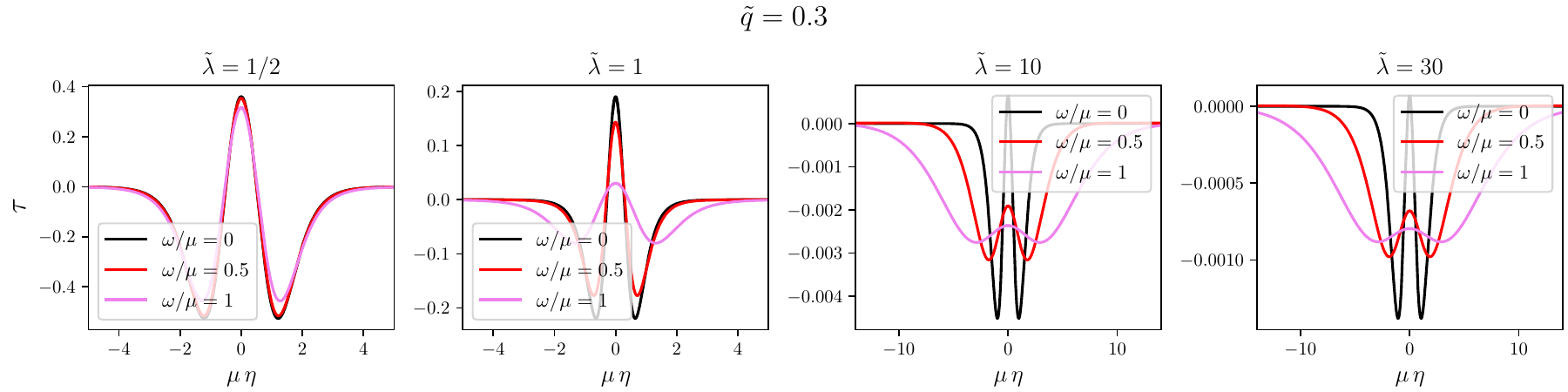}
            \includegraphics[scale=.5]{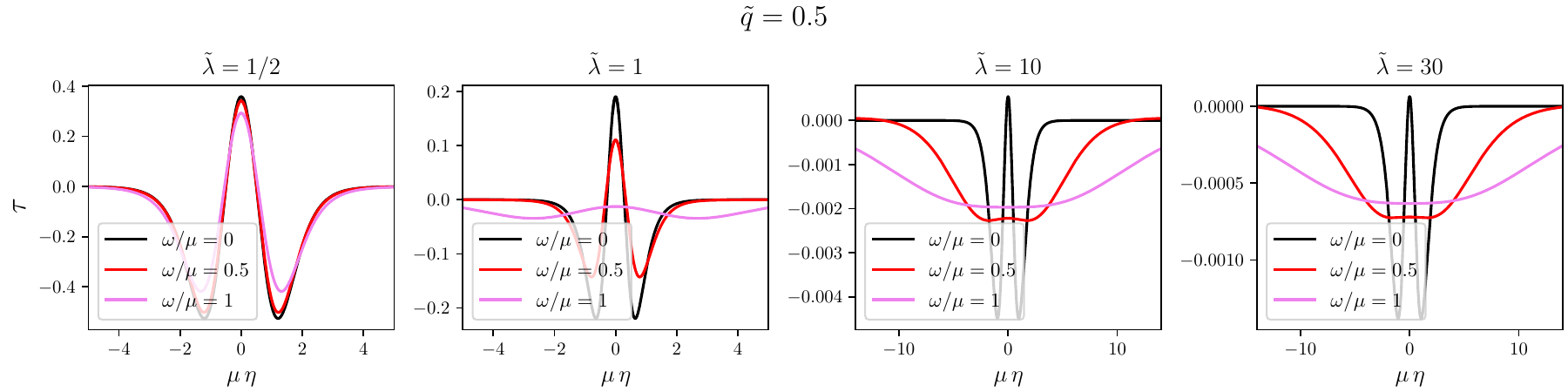}
		\par
	\end{centering}
\caption{The energy density $\tau$ as a function of $\eta$ for several wormhole solutions with $\tilde{\lambda}\in[0.5,30]$ and $\tilde{q}\in[0,\,0.5]$. The density energy  decreases its amplitude  slightly with $\tilde{\lambda}$ and the charge  $\tilde{q}$ for $\tilde{\lambda}\geq 0.5$ decreases the density energy profile, turning it negative in all the regions.} \label{fig:Es}
\end{figure}

The sources of the gravitational field are also helpful to clarify the structure of the wormhole as well as its global properties. In Fig.~\ref{fig:Es},
 we plot the energy density $\tau$ for several solutions with $\tilde{\lambda}\in[0.5,\,30]$, frequency $\omega/\mu\in \left[ 0, 1\right]$ and charge $\tilde{q}\in \left[ 0, 0.5\right]$. The rows of panels increase the value of the $\lambda$-parameter in the interval $\tilde{\lambda}\in[0.5,\, 30]$ from left to right and, from top to bottom the panel columns increase the $q$-parameter. When $\tilde{\lambda} \rightarrow 0$, the energy distribution has qualitatively the same profile for all $\omega/\mu$ values, which already indicates that the boson frequency becomes irrelevant for wormholes with $\tilde{\lambda}\rightarrow  0$, (recall that for $\tilde{\lambda}=0$, there are no wormhole solutions). Moreover, for small values of $\tilde{\lambda}$ the role of the charge is also negligible, whereas for larger values of $\tilde{\lambda}$ the role of the charge in the energy density becomes important, flattening the value of the energy density at the center.

\begin{figure}[ht]
	\begin{centering}	
            \includegraphics[scale=0.5]{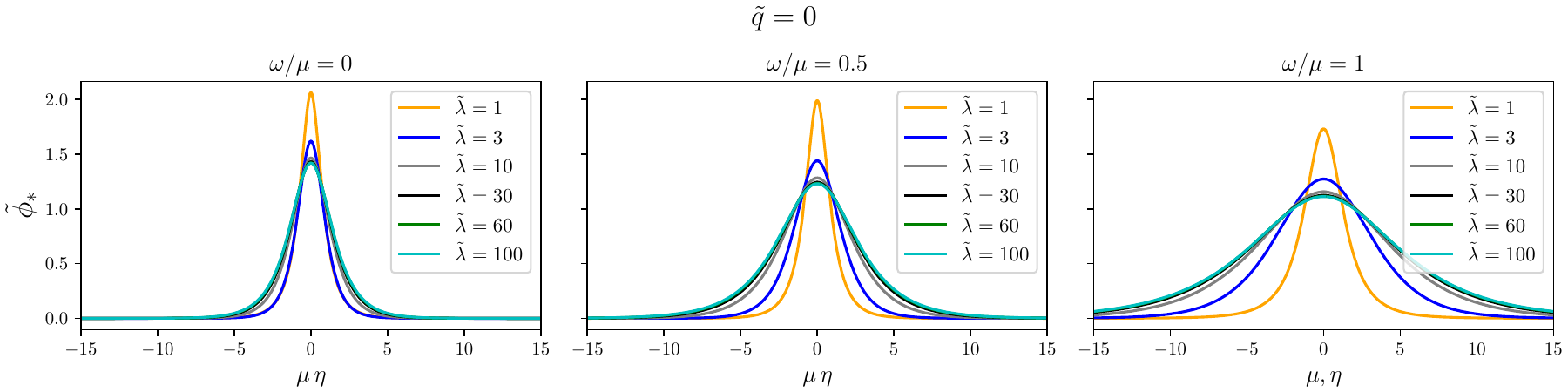}
            \includegraphics[scale=0.5]{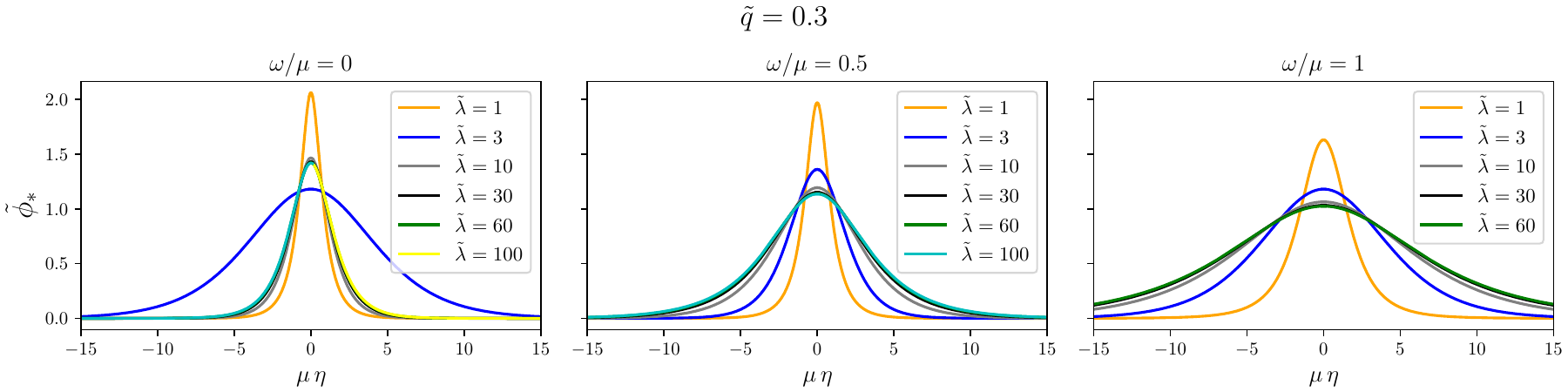}
		\par
	\end{centering}
	\caption{  In the figure, we present some numerical  solutions for the scalar field scaled by the factor $\sqrt{\lambda}$ and different values of $q$. The solutions are  similar for large $\sqrt{\lambda}$ whereas for small values of $\sqrt{\lambda}$ the solutions have very different profiles.} 
	\label{fig:phis_*}
\end{figure}

In Fig.~\ref{fig:phis_*} some solutions for  the scalar field $\phi$ are presented, where $\phi$ is scaled by the factor  $\sqrt{\lambda}$. We can see that for large values of $\lambda$ the solutions present a similar behaviour, indicating a similarity in the behavior of such solutions under the scaling  $\phi\to\sqrt{\lambda}\ \phi$ for  $\lambda\to\infty$. 

In this way, we have seen that for large $\lambda$, the particle number $\mathcal{N}$ and the throat radius ${\cal G}$, increase with $\lambda$, the mass $M$ depends linearly on $\lambda$ and finally we see that the $\sqrt{\lambda}\ \phi$ solutions tend to have a very similar profile for large $\lambda$.  These results suggest us to define  the scaled quantities $\phi_{*}:=\sqrt{\lambda}\ \phi$, $\eta_{0*}:=\eta_{0}/\lambda$, $\mathcal{N_{*}}:=\mathcal{N}/\lambda$ and $M_{*}:=M/\lambda$ in order to study the common behavior of the solutions with a fixed frequency $\omega$ and charge $q$  when $\lambda\to\infty$. This will be done in the following section.

\begin{figure}
	\begin{centering}
            \includegraphics[scale=.5]{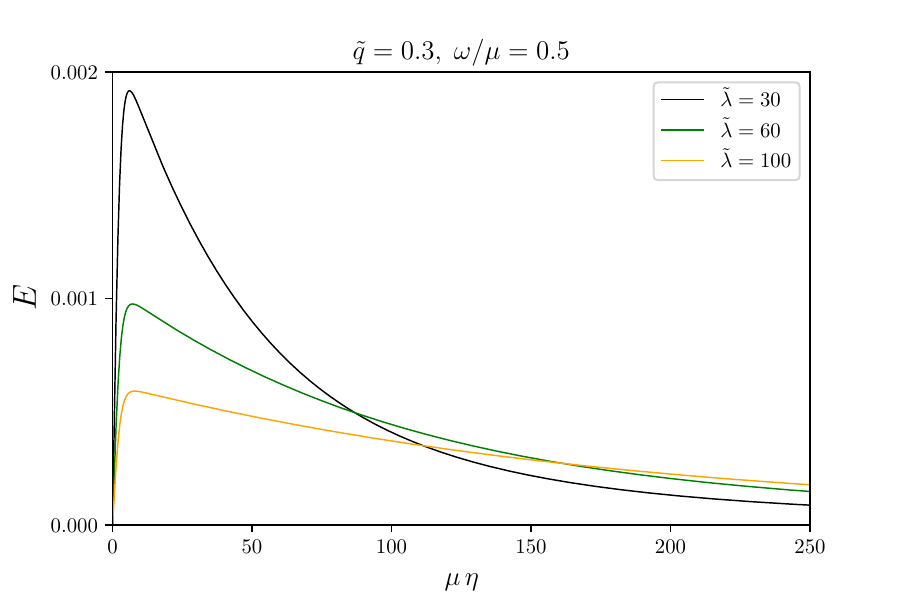}
            \includegraphics[scale=.5]{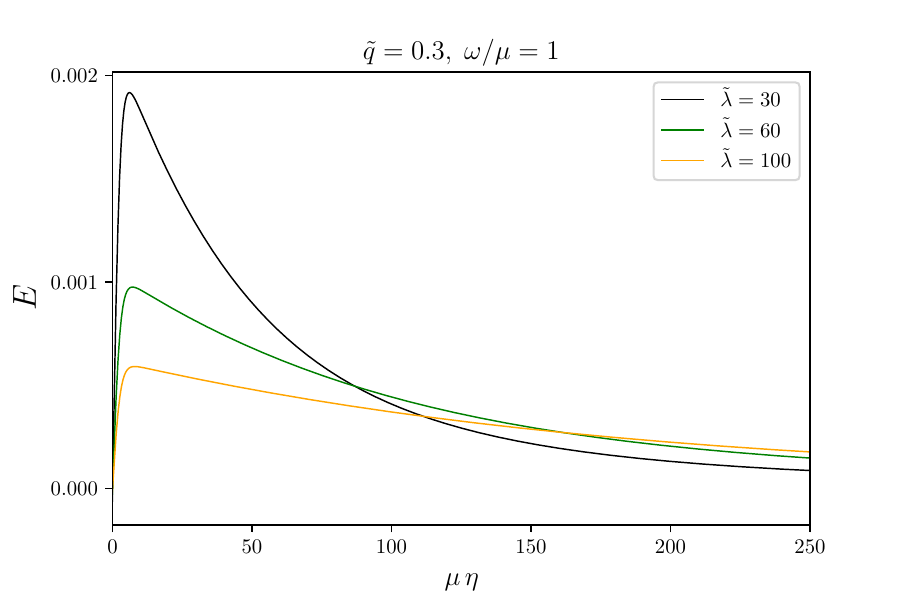}
            \includegraphics[scale=.5]{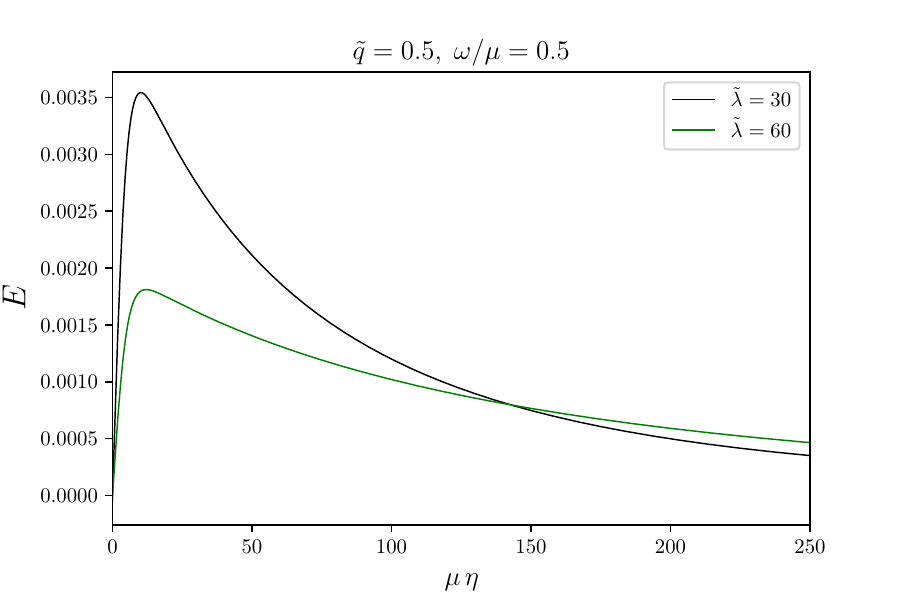}
            \includegraphics[scale=.5]{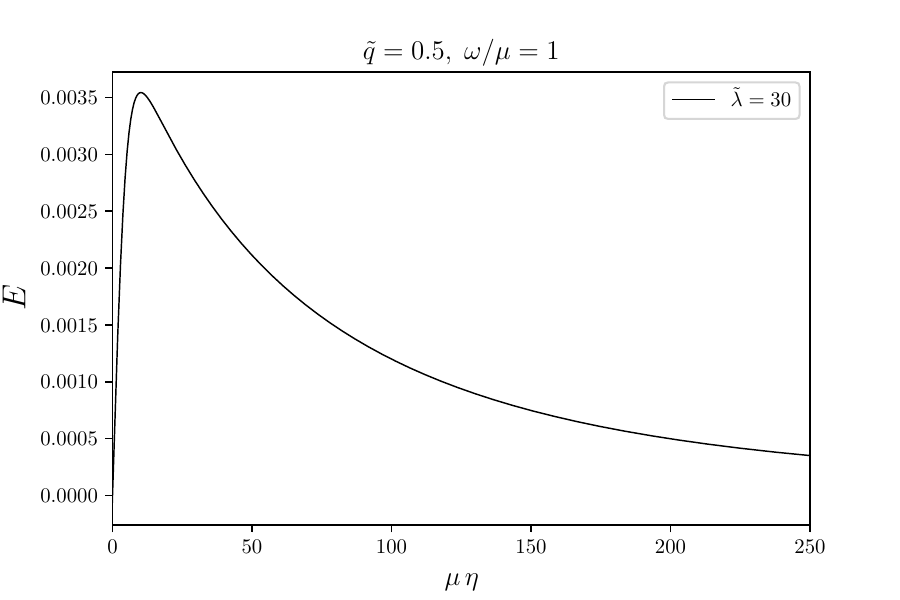}
		\par
	\end{centering}
	\caption{Electric field $E$ ($-\nabla V$) as a function of $\eta$ for  $\omega/\mu=0.5$ (left) and $\omega/\mu=1$ (right), with charges $\tilde{q}= 0.3$ (upp) and $\tilde{q}= 0.5$ (down).   The Electric field increases with $\tilde{q}$. In the case of large $\lambda$ it is difficult to obtain solutions near $\tilde{q}=0.5$, therefore  solutions for very large $\tilde{\lambda}$ are possible only for small values of $\tilde{q}$. } 
	\label{fig:Vs}
\end{figure}

We conclude this section presenting, in Fig.~\ref{fig:Vs}, the electric field profiles $E=-\nabla V$ for some values of the parameters $\lambda$,  $\omega$ and $q$. 
For our numerical implementation, the repulsive effect of the parameter $q$ allows us to obtain solutions up to a value $\tilde{q}\sim 0.5$, beyond which solutions become increasingly larger in size, this fact might also point to the existence of a critical value of the charge of the wormhole, already mentioned above. The $\omega$ parameter, in addition, also affects the difficulty of obtaining solutions even with modest values of $q$. In some of our results, we will be able to plot solutions with a fixed value of $q$ and several values of the frequency up to a certain limit  $\omega/\mu<1$, as can be seen in this figure. In the next section, we will put together our findings and derive a possible explanation for such behavior.
%

\subsection{Behavior for large \texorpdfstring{$\lambda$}{l}}
\label{large_lambda}

Based on the results obtained with the code presented in the above section, and following the scaling analysis for boson stars given by Colpi et al. for large values of $\lambda$ \cite{Colpi:1986ye}, we now consider the electric wormhole solutions when $\lambda \gg 1$. An analysis of the behavior of the spacetime geometry and the radial profile of the scalar field when the self-interaction is very large makes evident the scaling of the different functions and parameters of the solution. For example, it has already been shown in Fig.~\ref{fig:phis_*} that the solutions converge to a $\lambda$-independent profile $\phi_{*}=\sqrt{\lambda} \, \phi$ as $\lambda$ increases. On the other hand, there is a clear linear scaling for the throat parameter, $\eta_{0*} = \eta_0 / \lambda$, while the metric coefficients $\Psi, N$ and the electric potential converge to constant functions, $\Psi_c, N_c$ and $V_c$ respectively, in a central region whose size (in $\eta$) increases also linearly with $\lambda$.
In this way, our results show that not only the scalar field becomes independent of $\lambda$ for large values of the self-interaction but the entire solution.
With this result in mind, it is possible to find analytical solutions for an electric wormhole in the case of $\lambda \gg 1$, the neutral wormhole included.

Let us start by scaling the constraint equation (\ref{eq:constriction}) for the wormhole throat obtaining:
\begin{equation} \label{eq:const}
    4\pi\Psi_c^4\eta_{0*}^2 \phi_{c*}^2 = \frac{1}{\lambda \left(-\mu^2+\frac{\phi_{c*}^2}{2} + \frac{\left(V_c\,q+\omega \right)^2}{N_c^2}\right)}\ ,
\end{equation}
where, since the quantities $\phi_{c*}$, $N_c$ and the product $\Psi_c^4\eta_{0*}^2 \phi_{c*}^2$ are finite for $\lambda\to\infty$, then the right hand side of Eq.~(\ref{eq:const}) must  have the following dependence in order to maintain consistency in the constraint equation:
\begin{equation} \label{eq:con}
    \mu^2-\frac{\phi_{c*}^2}{2} - \frac{\left(V_c\,q+\omega \right)^2}{N_c^2} \sim  \frac{1}{\lambda }+ \mathcal{O} (\lambda^{-2})\ .
\end{equation}

Next, we take the pointwise limit $\lambda\to\infty$, keeping $\eta$ fixed in Eq.~\eqref{eq:con}, so that defining the following limit quantities with subscript $\infty$,
\begin{equation}\label{eq:soleta0}
   N_{\lambda_{\infty}}= N|_{\lambda\to\infty}, \quad \Psi_{\lambda_{\infty}}=\Psi|_{\lambda\to\infty}, \quad \phi_{*\lambda_{\infty}}=\phi_{*}|_{\lambda\to\infty},\quad V_{\lambda_{\infty}}= V|_{\lambda\to\infty} \ , 
\end{equation}
allows to express the value of the scalar field at the wormhole throat in the limit $\lambda \rightarrow \infty$ as,
\begin{equation}\label{eq:phi0_lambda}
\phi_0 := \phi_{*\lambda_{\infty}}(\eta=0)=\sqrt{2 \left(\mu^2 - \frac{\left(\omega+q\,V_0\right
    )^2}{N_0^2}\right)} \ ,
\end{equation} 
where $V_0 := V_{\lambda_{\infty}}(\eta=0)$ and $N_0 := N_{\lambda_{\infty}}(\eta=0)$. This equation implies that whenever $\omega+q\,V_{\lambda_{\infty}}\gg1$, the value of the lapse must take considerably larger values at the origin  to maintain Eq.~\eqref{eq:phi0_lambda} real, which explains why it is more difficult to obtain the numerical solution for cases with large $q$ and $\omega$. This can be corroborated by looking at the first two columns of Tables \ref{tab:Tab2} and \ref{tab:Tab3}, which will be properly presented below.

Now, inserting the scaled scalar field profile $\phi_*$ and the scaled $\eta_{0*}$ in the  Einstein-Klein-Gordon-Maxwell system and taking the limit $\lambda \rightarrow \infty$ (with $\eta$ held fixed and assuming $|\eta/\eta_0|\ll1$), the following set of differential equations is obtained:
\begin{eqnarray} \label{eq:set_whe_lambda}
    &&\frac{d^2}{d \eta^2}\Psi_{\lambda_{\infty}} =0\ , \\
	&&\frac{d^2}{d \eta^2} N_{\lambda_{\infty}}=0\ , \\
	 &&\frac{d^2}{d \eta^2} V_{\lambda_{\infty}} =0\ ,\\ \label{eq:phi*}
	&&\frac{d^2}{d \eta^2}{\phi_{*\lambda_{\infty}}} -{\Psi_{\lambda_{\infty}}}^4\left(\mu^2-\,{\phi_{*{\lambda_{\infty}}}}^2-\left(\frac{qV_{\lambda_{\infty}}+\omega}{N_{\lambda_{\infty}}}\right)^2\right)\phi_{*{\lambda_{\infty}}}=0\ .
\end{eqnarray}
Whose solution is valid in the interval $-\lambda\eta_{0*}\ll\eta\ll\lambda\eta_{0*}$.

Imposing symmetric solutions at the throat (Eq.~(\ref{eq:regularity_eta})) and using the definitions given in Eq.~(\ref{eq:soleta0}), the system \eqref{eq:set_whe_lambda} has the solution,
\begin{eqnarray}\label{eq:sol}
 && N_{\lambda_{\infty}}(\eta)= N_0, \quad \Psi_{\lambda_{\infty}}(\eta)=\Psi_0, \quad V_{\lambda_{\infty}}(\eta)= V_0\ ;\\
&&\phi_{*\lambda_{\infty}} (\eta)=\phi_0\  \rm{sech} \left(\frac{{\Psi_0}^2 
 \phi_{0}}{\sqrt{2}} \eta \right) \label{eq:phi_lambda}
 \ ,
\end{eqnarray}
with $\phi_0$ constrained by Eq.~(\ref{eq:phi0_lambda}). Given $q$ and $\omega$, the numbers $N_{0}, \ \Psi_{0},\ V_{0}$ and $\phi_{0}$ are estimated by interpolation from solutions with high values of $\lambda$. In Table \ref{tab:Tab2} we present some of these numerical values for the $\omega/\mu=0.5$ case. In Table \ref{tab:Tab3} we report the same quantities but now restricting to the $q=0$ case.
\begin{table}
\caption{\label{tab:Tab2}  
Limit quantities $N_{\lambda_{\infty}}$, $\Psi_{\lambda_{\infty}}$, $\phi_{\lambda_{\infty}}$ and $V_{\lambda_{\infty}}$ for $\omega/\mu=0.5$ and different values of the charge $q$. These quantities are related to the boundary conditions at the throat for $\lambda \to \infty$ according to the relation Eq.~(\ref{eq:soleta0}) and Eq.~(\ref{eq:phi0_lambda}). 
}
\begin{tabular}{cccccccccccc}
\hline 
$\tilde{q}$ &  & $N_{0}$ &  & $\Psi_{0}$ &  & $\phi_{0}$ &  & $V_{0}$ &  & $(\omega+qV_{0})/N_{0}$ & \tabularnewline
\hline 
\hline 
$0$ &  & $1$ &  & $0.7099$ &  & $1.225$ &  & $0$ &  & $0.5$ & \tabularnewline
$0.2$ &  & $1.0860$ &  & $0.6897$ &  & $1.193$ &  & $0.0882$ &  & $0.5418$ & \tabularnewline
$0.3$ &  & $1.2153$ &  & $0.6617$ &  & $1.1365$ &  & $0.1515$ &  & $0.5990$ & \tabularnewline
$0.5$ &  & $1.9767$ &  & $0.5504$ &  & $0.8926$ &  & $0.4146$ &  & $0.7787$ & \tabularnewline
\hline 
\end{tabular}
\end{table}
\begin{table}
\caption{\label{tab:Tab3}  
Limit quantities $N_0$, $\Psi_0$ and $\phi_0$  for $q=0$ ($V_{0}=0$) and different values of $\omega/\mu$.}
\begin{tabular}{cccccccccc}
\hline 
 $\omega/\mu$ &  & $N_0$ &  & $\Psi_0$ &  & $\phi_0$ &  & $\omega/N_0$ &  \tabularnewline
\hline 
\hline 
 $0$ & & $0.5820$ &  & $0.9001$ &  & $1.4189$ &  & $0$ & \tabularnewline
 $0.2$ & & $0.6895$ &  & $0.8427$ &  & $1.3551$ &  & $0.2901$ & \tabularnewline
 $0.5$ & & $1$ &  & $0.7099$ &  & $1.2250$ &  & $0.5$ & \tabularnewline
 $0.7$ & & $1.2401$ &  & $0.6347$ &  & $1.1707$ &  & $0.5645$ & \tabularnewline
 $1$ & & $1.6153$ &  & $0.5448$ &  & $1.1131$ &  & $0.6191$ & \tabularnewline
\hline 
\end{tabular}
\end{table}

\begin{figure}[ht]
	\begin{centering}
	\includegraphics[scale=.5]{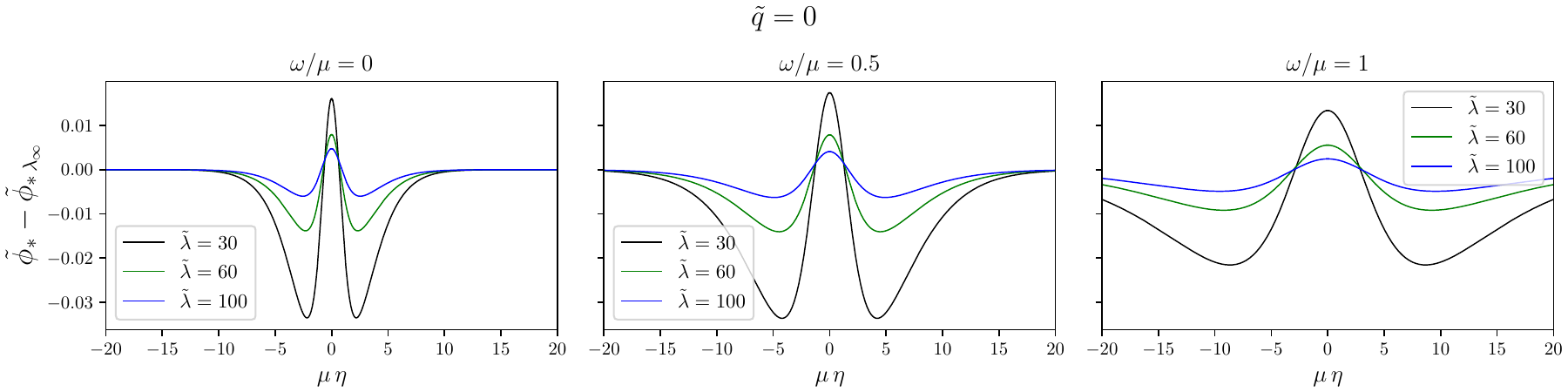}
        \includegraphics[scale=.5]{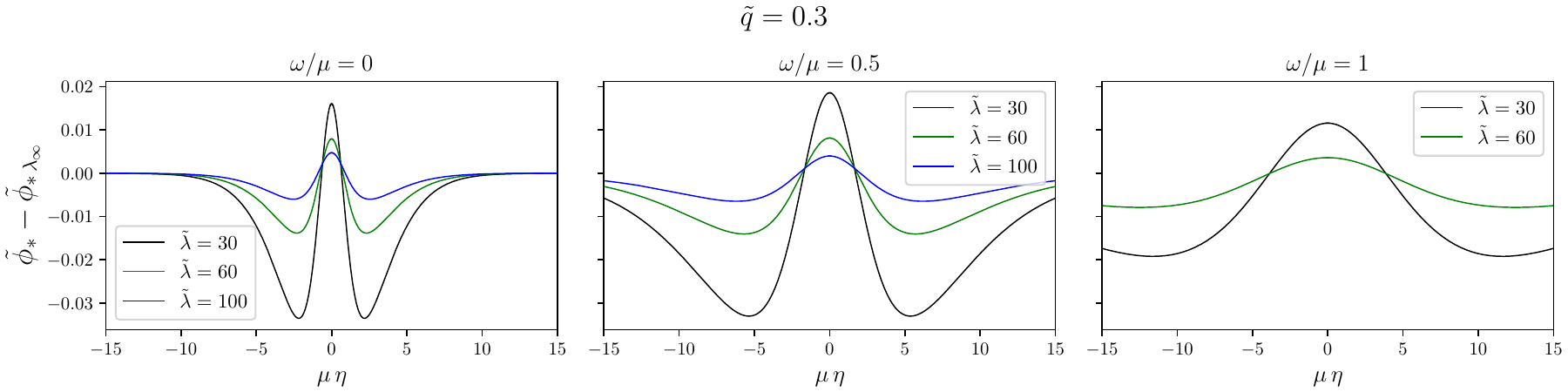}
		\par
	\end{centering}
	\caption{Difference between the numerical $\tilde{\phi}_{*}$ and the analytic scaled scalar field $\tilde{\phi}_{*\lambda_{\infty}}$, given in Eq.~(\ref{eq:phi_lambda}) and the parameters  given in the Table.~\ref{tab:Tab2}, as a function of the radial coordinate $\eta$   for large values of $\tilde{\lambda}$, for the cases with $\omega/\mu=0,\  0.5$ and $1$ for the frequency, and $\tilde{q}=0$, $\tilde{q}=0.3$ for the charge parameter. 
 } 
	\label{fig:phirescalado}
\end{figure}

In order to compare our analytic results for $\lambda\to \infty$ with numerical solutions, in Fig.~\ref{fig:phirescalado} we present the convergence of the numerical profiles $\phi_{*\lambda_{\infty}}$ to the analytic one, given  in Eq.~(\ref{eq:phi_lambda}), therefore concluding that the analytic expression gives a good approximation for these cases. We have also verified on the explored solutions, that the range of validity of the constant radial profiles \eqref{eq:sol} for the metric functions and the electric potential grows in size linearly with $\lambda$. 

Previously in the manuscript it has been described that the mass (and the number of particles) also have a linear scaling in the $\lambda\gg1$ case, and even more remarkable, the total mass goes to zero for the configuration with exactly $\omega=0.5\mu$ as $\lambda\to\infty$. The solutions obtained in Eq.~\eqref{eq:sol} and \eqref{eq:phi_lambda} provide an explanation of these properties. Using the scaled $\phi_{*\lambda_{\infty}}$ and $\eta_{0*}$ it is possible to show that the Komar mass in Eq.~(\ref{eq:M_Komar}) scales as $M_* = M/\lambda$ for  $ \lambda \gg 1$. Furthermore, an analytic expression for  $M_* $  can be obtained by noting that when $\lambda\gg1$ the square root of the determinant can be approximated as $\sqrt{\gamma}  \approx \lambda^{2}\Psi_{\lambda_{\infty}}^{6}\eta_{0*}^{2}\sin\theta$, and the integrand $N\left(T_{\  \mu}^{\mu}-2T_{\ t}^{t}\right)$ can be simplified to leading order in $1/\lambda$ as
\begin{equation}
N\left(T_{\  \mu}^{\mu}-2T_{\ t}^{t}\right)\approx 
\frac{2 N_{\lambda_{\infty}}\phi_{*\lambda_{\infty}}^{2}}{\lambda}\left(\frac{\mu^{2}}{2}-\frac{\phi_{*\lambda_{\infty}}^{2}}{4}-\frac{\left(qV_{\lambda_{\infty}}+\omega\right)^{2}}{N_{\lambda_{\infty}}^{2}}\right)\ .
\end{equation}
Thus, by inserting into the Komar mass expression (\ref{eq:M_Komar}), we obtain 
\begin{eqnarray}
M_{*} & = & 8\pi N_{0}^{}\,\Psi_{0}^{6}\,\eta_{0*}^{2}\left[-\frac{1}{4}\int_{0}^{\infty}\phi_{*\lambda_{\infty}}^{4}\,d\eta+\left(\frac{\mu^{2}}{2}-\frac{\left(\omega+qV_{0}\right)^{2}}{N_{0}^{2}}\right)\int_{0}^{\infty}\phi_{*\lambda_{\infty}}^{2}\,d\eta\right],\nonumber \\
 & = & \frac{8}{3}\pi N_{0}^{}\,\Psi_{0}^{4}\,\eta_{0*}^{2}\,\left(\mu^{2}-\frac{\left(\omega+qV_{0}\right)^{2}}{N_{0}^{2}}\right)^{1/2}\left[\mu^{2}-\frac{4\left(\omega+qV_{0}\right)^{2}}{N_{0}^{2}}\right]\ \label{M_roots} .
\end{eqnarray}
Here we have used the expression of the scalar field solution,  Eq.~(\ref{eq:phi_lambda}), which also imply that $\int_{0}^{\infty}\phi_{*\lambda_{\infty}}^{2}\,d\eta=\frac{2\phi_{0}}{\sqrt{2}\Psi_{0}^{2}},\,\,\int_{0}^{\infty}\phi_{*\lambda_{\infty}}^{4}\,d\eta=\frac{4\phi_{0}^{3}}{3\sqrt{2}\Psi_{0}^{2}}$.

As we were looking for, the above equation allows us to see that the mass of the system will be zero if and only if the condition $\omega=\omega_{\rm \rm zm}$ is satisfied, with
\begin{equation}
    \omega_{\rm \rm zm}=\frac{\mu N_{0}}{2}-  q\,V_{0}\ .
\end{equation} 

Substituting the values in Table \ref{tab:Tab2} for the solution $q=0$ and $\omega = 0.5\mu$ we see that it satisfies this condition and the same can be checked for configurations with $q>0$, $\lambda\gg1$ and whose $M$ is equal to zero. Moreover, it can be shown\footnote{
Considering the full expansion of the lapse function, after the constant term $N_{\lambda_\infty}$, consistent with the system (\ref{eq:set_electrowormhole}-\ref{eq:set_elec_wh_V}) (and considering the possible contributions of the expansions of the other fields) one must have $N=N_{\lambda_\infty}+\lambda^{-1}N_1(\eta)+\mathcal{O}(\lambda^{-2})$, with $N_1$ satisfying the differential equation $N_1''=4\pi N_{\lambda_\infty}\Psi_{\lambda_\infty}^4\phi_*^2(\mu^2-\phi_*^2/2-2\omega^2/N_{\lambda_\infty}^2)$ subject to the boundary condition $N_1'(\eta=0)=0$. Under this consideration, the solution is,
\begin{equation}
    N_1(\eta)= 4\pi N_{\lambda_\infty}\phi_{*\lambda_\infty}^2\left[\left(\frac{1}{3}-\kappa \right)\ln\left(\cosh\left(b\eta\right)\right)+\frac{1}{6}\mathrm{sech}^2\left(b\eta\right)\right]+k_1,
\end{equation}
with $\kappa=2\phi_{*\lambda_\infty}^{-2}\omega^2/N_{\lambda_\infty}$ and $b=\Psi_{\lambda_\infty}^2\phi_{*\lambda_\infty}/\sqrt{2}$. Recalling these solutions are valid in the domain $|\eta|<\lambda\eta_{0*}$, the full lapse function $N=N_{\lambda_\infty}+\lambda^{-1}N_1(\eta)+\mathcal{O}(\lambda^{-2})$ should match the exterior solution $N_{\eta\to\infty}=1+\kappa_1/\eta$ at a certain point $1\ll\eta_m<\lambda$, but it is precisely in the case of zero mass that $\kappa=1/3$, as obtained from Eq.~\eqref{eq:phi_lambda}, so the term $\lambda^{-1} N_1$ contributes negligibly (not so whenever $\kappa\neq1/3$). Furthermore, it can be argued that the $\mathcal{O}(\lambda^{-2})$ contributions are equally negligible at the matching point so that $N_{\lambda_\infty}$ should meet the boundary condition at infinity, $N_{\lambda_\infty}=1$, for the $q=0$ family of solutions if and only if $M_*=0$.
} that in the neutral case, the value of $N_{0}$ is exactly equal to $1$, thus explaining why it is precisely at $\omega=0.5\mu$ that the mass of the spacetime tends to zero as already noted in \cite{Dzhunushaliev:2008bq, Chew:2019lsa}. In general, such value is modified by the presence of the electromagnetic coupling $q$, a behavior that was hinted at in the previous plots of the mass.
We noticed from Tables \ref{tab:Tab2} and \ref{tab:Tab3} that the  quantities  $N_{0}$ and $V_{0}$ are monotonically increasing with the frequency and the charge, while $\phi_{0}$  and $\Psi_{0}$ decrease monotonically with the frequency and charge, therefore for each value of the frequency and charge, there corresponds a single value of $N_{0}$ such that the mass $M_{*}$ is equal to zero. 

Finally, using the same procedure it is possible to see that the number of particles $\mathcal{N}$ given by  Eq.~(\ref{eq:KomarN}) scale  as $\mathcal{N}_{*}=\mathcal{N}/\lambda$ for large $\lambda$. An analytic expression for $\mathcal{N}_{*}$ can also be derived and is given by  

\begin{eqnarray}
\mathcal{N}_{*} & = & 4\pi\, \,\eta_{0*}^{2}\left(\omega+qV_{0}\right)\frac{\Psi_{0}^{6}}{ N_{0} }\int_{0}^{\infty}\phi_{*\lambda_{\infty}}^{2}\,d\eta\ ,\nonumber \\
 & = & 8 \pi \,\eta_{0*}^{2} \left(\omega+qV_{0}\right)\, \frac{\Psi_{0}^{4}}{ N_{0} } \left(\mu^{2}-\frac{\left(\omega+qV_{0}\right)^{2}}{N_{0}^{2}}\right)^{1/2} ,
\end{eqnarray}
where we have used the current Eq.~(\ref{eq:current}) and the ansatz Eq.~(\ref{eq:ansatz_phi}) to obtain
$j^0=-\frac{{\phi}^2_{\lambda_{\infty}}}{N^2_{\lambda_{\infty}}}\,\left(V_{\lambda_{\infty}}\,q+\omega\right)$. 
%

\subsection{Particle motion} \label{motion}

The analysis of the motion of particles in the several types of spacetimes generated by the charged wormhole throws interesting facts that are useful in the understanding of the properties of such geometries.

The full action of a charged particle with mass $m$ and charge $e$ interacting with an electromagnetic field in General Relativity reads \cite{Padmanabhan:2010zzb},
\begin{equation}
    {\cal A} = -\int m d\tau + \int e A_\mu u^\mu d\tau + \int d^4 x\sqrt{-g} \left[\frac{1}{16\pi }\mathcal{R}- \frac{1}{4}F_{\mu\nu}F^{\mu\nu}\right]\ .
\end{equation}
Where $u^\mu$ is the four-velocity of the particle. Varying ${\cal A}$ with respect to the trajectory of the particle leads to the equations of motion of such particle (Lorentz force law),
\begin{equation}\label{eq:lorentz}
    m\ u^\mu\nabla_\mu u ^\nu = e\ {F^{\nu}}_\alpha u^\alpha\ .
\end{equation}
Now, let $K$ be a Killing vector field of the spacetime, then it can be shown that the Killing equation $\nabla_{(\mu} K_{\nu)} = 0$ does not imply $u^\mu\nabla_\mu (K_\nu u^\nu)=0$ as in the case of geodesics, but rather $u^\mu\nabla_\mu \left[K_\nu (m u^\nu + e A^\nu)\right]=0$ when Eq.~\eqref{eq:lorentz} is used and it is assumed that the electromagnetic field is consistent with the symmetry associated with $K$ \cite{Wald:1984rg} (as is the case for the electric wormhole solution) . This means that, the quantities
\begin{equation}
    K_\nu (m u^\nu + eA^\nu)\ ,
\end{equation}
are constant along the world line of the charged particle. Returning to the wormhole spacetime, the timelike Killing field $\xi$ and the axial Killing field $\psi=\partial_\varphi$ imply the existence of a conserved energy $\mathcal{E}:=-\xi_\nu (m u^\nu + eA^\nu)$ and a conserved (azimuthal) angular momentum $L:=\psi_\nu (m u^\nu + eA^\nu)$, which in the coordinates described in Eq.~\eqref{eq:eleS} we obtain $\mathcal{E}=mN^2u^t-eV$ and $L=m \Psi^4(\eta^2 + \eta_0^2)\ \sin^2\theta\  u^\varphi$.
Since both the gravitational and electromagnetic fields are spherical we may restrict to the equatorial motion $\theta = \pi /2 $ without loss of generality, so the normalization of the four-velocity allows us to obtain a simple equation for the radial motion:
\begin{equation}
     m^2N^2\Psi^4\left(\frac{\partial\eta}{\partial\tau}\right)^2 + m^2 N^2 - \left(\mathcal{E} + eV\right)^2 + \frac{N^2}{\Psi^4}\frac{L^2}{\eta^2+{\eta_0}^2}=0\ ,  
\end{equation}
whose solutions allow to acquire a better understanding of the properties of the wormholes determined by the parameters. Now, we can define an effective potential $U_{\rm eff}$ as the minimum allowed value of $\mathcal{E}$ at a given $\eta$, \textit{i.e.},
\begin{equation}
     U_{\rm eff}(\eta) = - eV(\eta) + \sqrt{m^2 N(\eta)^2 + \frac{N(\eta)^2}{\Psi(\eta)^4}\frac{L^2}{\eta^2+{\eta_0}^2}}\ .  
\end{equation}

Given $\mathcal{E}$ and $L$, then, the allowed regions for the movement of the particle are given by those values of $\eta$ such that $U_{\rm eff}(\eta) \leq \mathcal{E}$. 

\begin{figure}[h]
	\begin{centering}
		\includegraphics[scale=.5]{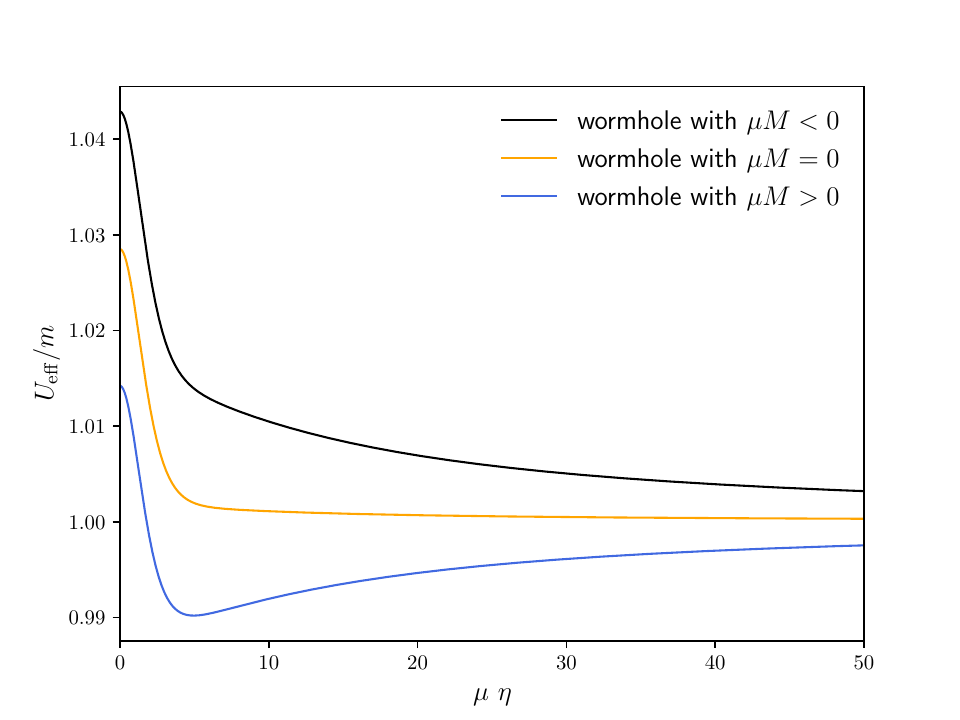}
        \includegraphics[scale=.5]{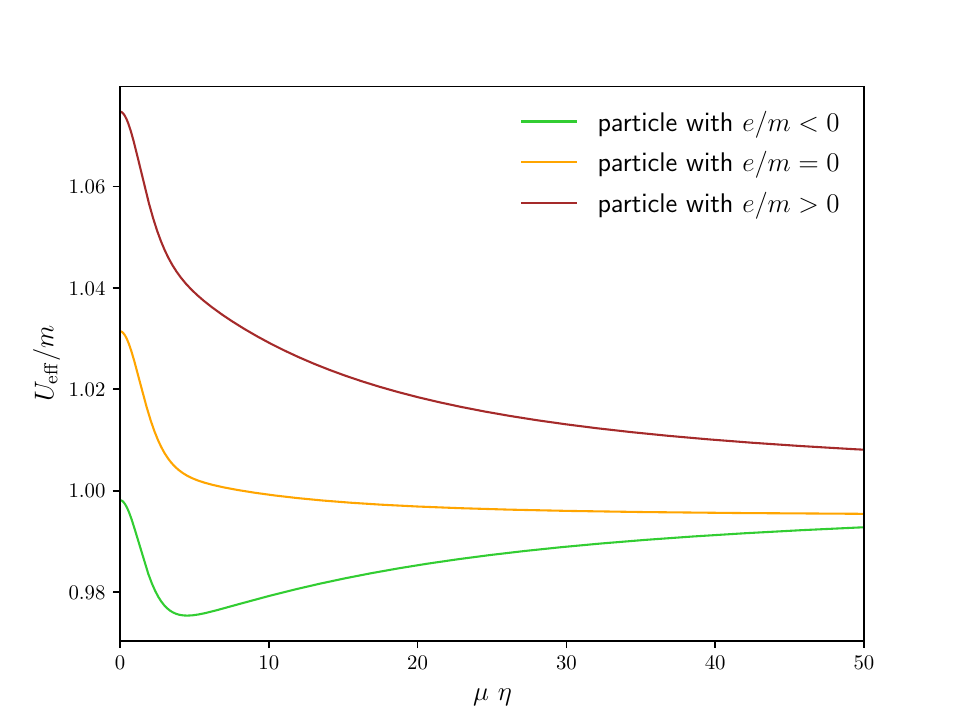}
		\par
	\end{centering}
	\caption{Effective potential $U_{\rm eff}$ for a particle with angular momentum $L/m=0.1$. Left panel: Neutral particle moving in an electric wormhole with $\tilde{\lambda} =  30$ and $\tilde{q} = 0.1 $ and different values of $\omega/\mu$ such that the total mass of the spacetime is positive zero and negative ($\omega/\mu = 0.475$, $0.4875$ and $0.5$ respectively). Right panel: Charged particles with $e/m = 1$, $0$, $-1$ moving in the same $M = 0$ wormhole as in the left panel.}
	\label{fig:Ueff}
\end{figure}

In order to illustrate the motion of particles around the spacetime of an electrically charged wormhole, some effective potentials are shown in Fig.~\ref{fig:Ueff}. In the left panel, we use three different spacetimes with $M>0$, $M<0$ and $M=0$ such that a neutral particle with angular momentum of $L/m=0.1$ will fall to the wormhole when the mass is positive, be repelled when the mass is negative and continuing in an almost straight line when $M=0$ and is far away from the wormhole throat. In the right panel of Fig.~\ref{fig:Ueff}, we fix the spacetime to be one with $M=0$ and give certain values for the charge of the particle, then we can see that the repulsive and attractive interaction towards the wormhole can be obtained through the electromagnetic field. 

\section{Conclusions}
\label{Sec:Conclusions}

We have derived and solved the Einstein-Maxwell-Klein Gordon system for the case of an exotic and massive scalar field with a self-interaction term, minimally coupled to the electromagnetic field. Imposing appropriate conditions in the boundaries, we obtain an asymptotically flat charged wormhole and analyze the solutions. We obtained solutions with the already known regions of positive energy density near the throat, followed by regions of negative density, obtaining configurations where the total mass of the system can be positive, negative or even zero, depending on the values of the parameters of the system. We obtain that the electric charge affects the morphology of the wormhole and plays an important role in the determination of such total mass and also in the total particle number of the system. The motion of particles in the regions far from such wormholes is different depending on the total mass of the system and  the charge.

Our analysis suggested that, for large values of the self-interaction parameter, $\lambda$, the solution for the scalar field tended to have similar behavior. We explored this fact and were able to obtain an analytic expression for the scalar field that reproduced its behavior in this case of the large self-interaction term, a fact that we proved by comparing the actual numerical solutions in this case with the analytic expressions, obtaining excellent agreement. This allowed also a better understanding of the role of the parameters in the determination of the wormhole properties, namely the total mass of the configuration and the particle number, as a function of the system parameters, $\mu$, $\omega$, $q$, $\eta_{0*}$, and the electric potential and metric coefficients evaluated in the case of large $\lambda$. It is interesting to notice that the charge and electric field play an important role not only in determining the value of the total mass but also in the fact that it is modified the case when such total mass is zero.

Another fact that we want to underline is that our numerical experiments allow us to present the conjecture that the charge can not play the role of the self-interaction constant regarding the existence of (equilibrium) solutions for the massive scalar field; that is, $\lambda$ has to be non zero in order to have a wormhole, even for the case of an electric wormhole. 

Also, even though there is no  equation which hints at the existence of a critical charge, a maximum value for the charge, the numerical experiments show that it is increasingly difficult to find solutions for $\tilde{q}$ larger than $0.5$, which might indicate that there is a critical value for the charge beyond which there are no equilibrium solutions. 

Finally, we presented the motion of particles, charged or neutral, obtaining an expected but interesting  behavior in terms of the total mass. Indeed, in the region where the total enclosed mass is already constant, the particle is attracted towards the wormhole when such mass is positive, repelled away from it, when the total mass is negative, and moves as if there were no wormhole at all when the total mass is zero. The $Q\neq0$ and $M=0$ situation is of importance because although the electromagnetic field contributes to the sources of the Einstein equations, the full system solution (electromagnetic field + scalar field) is such that a charged particle, far from the throat, would be able to sense the presence of the wormhole, whereas a neutral particle would not. It has been possible to construct solutions in which unlike the Kerr-Newman family hole one has $Q>M$ without implying the existence of a naked singularity. 

The ideas and procedure described in this work can be adapted to the case when the electromagnetic field has an azimuthal component, instead of a temporal one, obtaining a magnetic wormhole. Such work is currently under development.


\acknowledgments
This work was partially supported by the CONACyT Network Project No. 376127 ``Sombras, lentes y ondas gravitatorias generadas por objetos compactos astrof\'\i sicos``.
ML and VJ acknowledge financial support from CONACyT 
graduate grant program.  DMC acknowledge financial support from CONACyT through a postdoctoral research grant. DN acknowledge the sabbatical support given by the Programa de Apoyos para la Superaci\'on del Personal Acad\'emico de la Direcci\'on General de Asuntos del Personal Acad\'emico de la Universidad Nacional Aut\'onoma de M\'exico in the elaboration of the present work.

\appendix

\section{Geometric scalars}

The following expressions for the geometric scalars will be useful to characterize the solutions. The Ricci scalar:

\begin{equation}
 \mathcal{R}=-\frac{2}{\Psi^4}\,\left(\frac{\Delta_3\,N}{N} + 4\,\frac{\Delta_3\,\Psi}{\Psi} + 2\,\frac{\,N' \,\Psi'}{N\, \Psi}+ \frac{{\eta_0}^2}{\left(\eta^2 + {\eta_0}^2\right)^2}\right).
\end{equation}
The Weyl scalar
\begin{equation}
 W=\frac{4}{3\,\Psi^8}\,\left(\frac{N''}{N} - \frac{\eta}{\eta^2+{\eta_0}^2}\,\frac{N'}{N} -2\left( \frac{\Psi''}{\Psi} - \frac{\eta}{\eta^2+{\eta_0}^2}\,\frac{\Psi'}{\Psi}-3\,\left(\frac{\Psi'}{\Psi}\right)^2\right) - 4\,\frac{\,N' \,\Psi'}{N\, \Psi} - \frac{2\,{\eta_0}^2}{\left(\eta^2 + {\eta_0}^2\right)^2}\right)^2,
\end{equation}

and the Kretschmann scalar:
\begin{eqnarray}
K &=& \frac{4}{\Psi^8}\,\left(\frac{N''}{N}\left(\frac{N''}{N} - 4\,\frac{ N'\, \Psi'}{N 
\, \Psi}\right)  + \frac{8 \Psi''}{\Psi} \left(\frac{\Psi''}{\Psi} - \left(
2\left(\frac{\Psi'}{\Psi}\right)^2 - \frac{2\,\eta}{\eta^2+{\eta_0}^2}\, \frac{\Psi'}{\Psi}
- \frac{{\eta_0}^2}{\left(\eta^2+{\eta_0}^2\right)^2}\right)\right) \right. \nonumber \\
 && \left. 
 + 4\,\left(\frac{\Psi'}{\Psi}\right)^2\,\left(6\,\left(\frac{\Psi'}{\Psi}\right)^2 + 4\,\frac{\eta}{\eta^2+ {\eta_0}^2}\,\frac{\Psi'}{\Psi} + \frac{2\,\left(3\,\eta^2-2\,{\eta_0}^2\right)}{\left(\eta^2+{\eta_0}^2\right)^2} + 3\,\left(\frac{N'}{N}\right)^2\right)
 \right. \nonumber \\
 && \left. 
 + \frac{2}{\eta^2+{\eta_0}^2}\,\left(\frac{N'}{N}\right)^2\,\left(4\,\eta\,\frac{\Psi'}{\Psi} + \frac{\eta^2}{\left(\eta^2+{\eta_0}^2\right)^2}\right) + \frac{3\,{\eta_0}^2}{\left(\eta^2+{\eta_0}^2\right)^2}\right)\ .
\end{eqnarray}

\begin{figure}[H]
	\begin{centering}
        \includegraphics[scale=0.6]{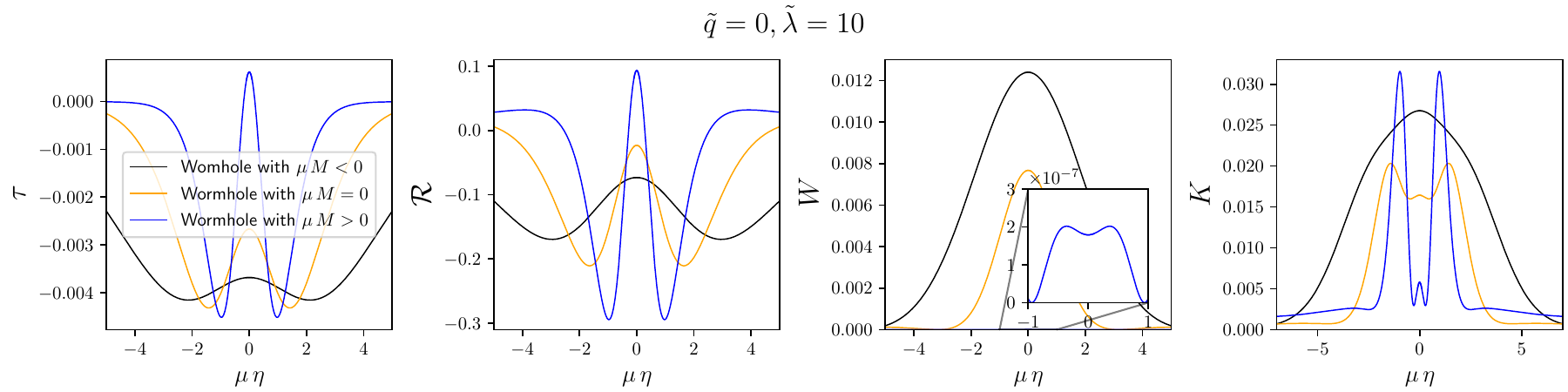}
        \includegraphics[scale=0.6]{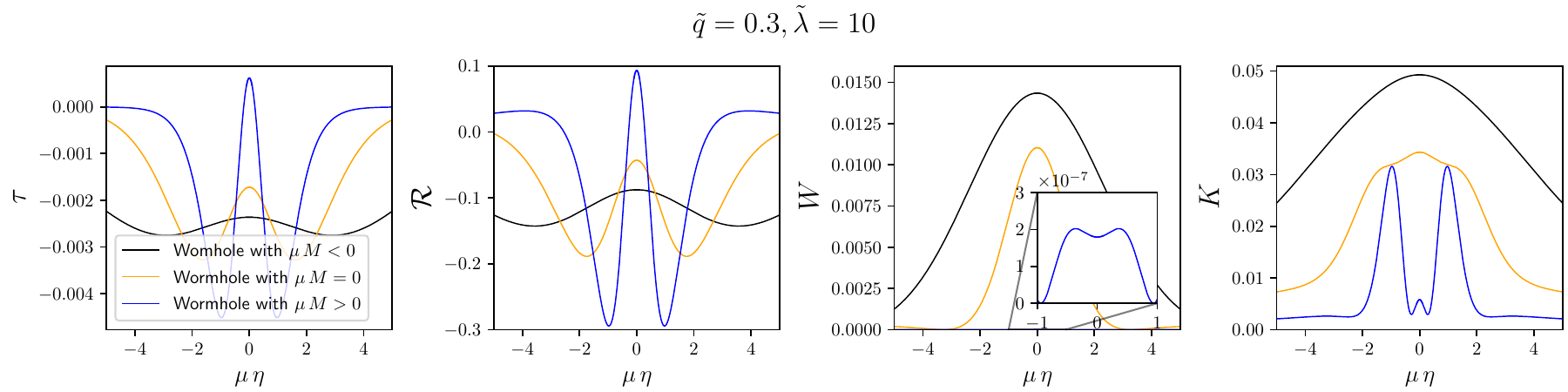}
		\par
	\end{centering}
	\caption{Ricci $\mathcal{R}$, Weyl $W$, Kretschmann $K$ scalars and density energy $\tau$ for  wormholes with $\tilde{\lambda}=10$, $\tilde{q} = \{0,\;0.3\}$ and   $\omega/\mu=\{0,\,\omega_{zm}/\mu,1\}$, such that the total mass of the spacetime is positive, zero and negative, respectively. We also plot the energy density $\tau$ for the same cases.} 
	\label{fig:scalars}
\end{figure}

In Fig.~\ref{fig:scalars}, we present some plots of these geometric scalars which might also help to better understand the role of the parameters in the wormhole configuration. Recall that the scalar of curvature $\mathcal{R}$ is proportional to the trace of the stress energy tensor, $\mathcal{R}=-8\,\pi\,\left(\tau + S\right)$, and in the figure we present both to stress the importance of the trace $S$ in determining the proportional changes for the different cases of the total mass.

It is remarkable the change on the geometry depending on the total mass of the system. Such influence is mainly showed in the Weyl scalar in which maximum for the zero mass case is almost half the one of the negative mass case and for the positive mass case, the profile of the Weyl scalar is more than four orders in magnitude smaller than in the other cases. These solutions invite for a deeper understanding on the properties of the geometric scalars depending on the matter-energy presented in the space-time. Such work will be done elsewhere.


\bibliography{ref}
\end{document}